\documentclass[%
  conference,%
  10pt,%
]{IEEEtran}

\usepackage{cite}
\usepackage{amsmath,amssymb,amsfonts}
\allowdisplaybreaks

\usepackage{paralist}

\usepackage[UKenglish]{babel}

\usepackage{booktabs}
\usepackage{url}

\usepackage{graphicx}
\graphicspath{{./images/}}
\usepackage{placeins}
\usepackage{subcaption}
\usepackage[justification=centering]{caption}
\usepackage{enumitem}
\usepackage{balance}
\usepackage{booktabs}  %
\usepackage{colortbl}

\usepackage{pgfplotstable}
\DeclareMathVersion{tabularbold}
\SetSymbolFont{operators}{tabularbold}{OT1}{cmr}{b}{n}

\usepackage{multirow}

\usepackage{inconsolata}

\usepackage{hyperref}

\usepackage[%
  group-minimum-digits=4,%
  list-final-separator={, and },%
  add-integer-zero=false,%
  free-standing-units,%
  binary-units,%
  detect-weight=true,%
  detect-inline-weight=math,%
]{siunitx}

\usepackage{microtype}

\usepackage[colorinlistoftodos,prependcaption,textsize=tiny
]{todonotes}

\usepackage{xspace}

\usepackage[capitalise]{cleveref}

\crefformat{footnote}{#2\footnotemark[#1]#3}

\usepackage{tabularx}

\newcommand{\multilinecell}[1]{%
    \begin{tabular}{@{}r@{}}#1\end{tabular}
}
\newcommand{\multilinecellL}[1]{%
    \begin{tabular}{@{}l@{}}#1\end{tabular}
}
\newcommand{\multilinecellC}[1]{%
    \begin{tabular}{@{}c@{}}#1\end{tabular}
}
\newcommand{\multilinecellR}[1]{%
    \begin{tabular}{@{}r@{}}#1\end{tabular}
}

\newcommand{\summary}[2]{
    \vspace{0.6em}
    \noindent
    \colorbox{gray!20}{%
        \parbox{.97\linewidth}{%
                \textbf{Summary \textit{(#1)}}
            #2
        }%
    }%
}%

\usepackage{csquotes}
\usepackage{csvsimple}
\usepackage{listings}
\usepackage{xcolor}

\definecolor{codegreen}{rgb}{0,0.6,0}
\definecolor{codegray}{rgb}{0.5,0.5,0.5}
\definecolor{codepurple}{rgb}{0.58,0,0.82}
\definecolor{backcolour}{rgb}{0.95,0.95,0.92}

\lstdefinestyle{mystyle}{
  backgroundcolor=\color{backcolour},
  commentstyle=\color{codegreen},
  keywordstyle=\color{magenta},
  numberstyle=\tiny\color{codegray},
  stringstyle=\color{codepurple},
  basicstyle=\ttfamily\footnotesize,
  breakatwhitespace=false,
  breaklines=true,
  breakindent=60pt,
  captionpos=b,
  keepspaces=true,
  numbers=left,
  numbersep=5pt,
  showspaces=false,
  showstringspaces=false,
  showtabs=false,
  tabsize=2,
  numberblanklines=true,
  showlines=true,
  xleftmargin=2em,
  framexleftmargin=2.2em
}

\lstMakeShortInline[language=Python]|

\let\origthelstnumber\thelstnumber
\makeatletter
\newcommand*\Suppressnumber{%
  \lst@AddToHook{OnNewLine}{%
    \let\thelstnumber\relax%
     \advance\c@lstnumber-\@ne\relax%
    }%
}

\newcommand*\Reactivatenumber[1]{%
  \setcounter{lstnumber}{\numexpr#1-1\relax}
  \lst@AddToHook{OnNewLine}{%
   \let\thelstnumber\origthelstnumber%
   \refstepcounter{lstnumber}
  }%
}

\newcommand{\ExamTarantulaAcImprovementOverFirstMeanNod}{\SI{28.1}{\percent}\xspace}
\newcommand{\ExamTarantulaAcImprovementOverFirstNumCasesGtzeroNod}{\num{59}\xspace}
\newcommand{\ExamTarantulaAcImprovementOverFirstNumCasesLtzeroNod}{\num{1}\xspace}
\newcommand{\ExamTarantulaAcImprovementOverFirstNumCasesEqzeroNod}{\num{21}\xspace}

\newcommand{\ExamTarantulaAcImprovementOverUnionMeanNod}{\SI{40.4}{\percent}\xspace}
\newcommand{\ExamTarantulaAcImprovementOverUnionNumCasesGtzeroNod}{\num{64}\xspace}
\newcommand{\ExamTarantulaAcImprovementOverUnionNumCasesLtzeroNod}{\num{1}\xspace}
\newcommand{\ExamTarantulaAcImprovementOverUnionNumCasesEqzeroNod}{\num{16}\xspace}

\newcommand{\ExamTarantulaAcImprovementOverIndividualMeanNod}{\SI{39.1}{\percent}\xspace}
\newcommand{\ExamTarantulaAcImprovementOverIndividualNumCasesGtzeroNod}{\num{62}\xspace}
\newcommand{\ExamTarantulaAcImprovementOverIndividualNumCasesLtzeroNod}{\num{1}\xspace}
\newcommand{\ExamTarantulaAcImprovementOverIndividualNumCasesEqzeroNod}{\num{18}\xspace}

\newcommand{\ExamOchiaiAcImprovementOverFirstMeanNod}{\SI{19.0}{\percent}\xspace}
\newcommand{\ExamOchiaiAcImprovementOverFirstNumCasesGtzeroNod}{\num{60}\xspace}
\newcommand{\ExamOchiaiAcImprovementOverFirstNumCasesLtzeroNod}{\num{0}\xspace}
\newcommand{\ExamOchiaiAcImprovementOverFirstNumCasesEqzeroNod}{\num{21}\xspace}

\newcommand{\ExamOchiaiAcImprovementOverUnionMeanNod}{\SI{31.7}{\percent}\xspace}
\newcommand{\ExamOchiaiAcImprovementOverUnionNumCasesGtzeroNod}{\num{65}\xspace}
\newcommand{\ExamOchiaiAcImprovementOverUnionNumCasesLtzeroNod}{\num{0}\xspace}
\newcommand{\ExamOchiaiAcImprovementOverUnionNumCasesEqzeroNod}{\num{16}\xspace}

\newcommand{\ExamOchiaiAcImprovementOverIndividualMeanNod}{\SI{8.4}{\percent}\xspace}
\newcommand{\ExamOchiaiAcImprovementOverIndividualNumCasesGtzeroNod}{\num{24}\xspace}
\newcommand{\ExamOchiaiAcImprovementOverIndividualNumCasesLtzeroNod}{\num{6}\xspace}
\newcommand{\ExamOchiaiAcImprovementOverIndividualNumCasesEqzeroNod}{\num{51}\xspace}

\newcommand{\ExamDstarAcImprovementOverFirstMeanNod}{\SI{18.7}{\percent}\xspace}
\newcommand{\ExamDstarAcImprovementOverFirstNumCasesGtzeroNod}{\num{60}\xspace}
\newcommand{\ExamDstarAcImprovementOverFirstNumCasesLtzeroNod}{\num{0}\xspace}
\newcommand{\ExamDstarAcImprovementOverFirstNumCasesEqzeroNod}{\num{21}\xspace}

\newcommand{\ExamDstarAcImprovementOverUnionMeanNod}{\SI{31.5}{\percent}\xspace}
\newcommand{\ExamDstarAcImprovementOverUnionNumCasesGtzeroNod}{\num{65}\xspace}
\newcommand{\ExamDstarAcImprovementOverUnionNumCasesLtzeroNod}{\num{0}\xspace}
\newcommand{\ExamDstarAcImprovementOverUnionNumCasesEqzeroNod}{\num{16}\xspace}

\newcommand{\ExamDstarAcImprovementOverIndividualMeanNod}{\SI{8.0}{\percent}\xspace}
\newcommand{\ExamDstarAcImprovementOverIndividualNumCasesGtzeroNod}{\num{24}\xspace}
\newcommand{\ExamDstarAcImprovementOverIndividualNumCasesLtzeroNod}{\num{6}\xspace}
\newcommand{\ExamDstarAcImprovementOverIndividualNumCasesEqzeroNod}{\num{51}\xspace}

\newcommand{\ExamOptwoAcImprovementOverFirstMeanNod}{\SI{17.0}{\percent}\xspace}
\newcommand{\ExamOptwoAcImprovementOverFirstNumCasesGtzeroNod}{\num{57}\xspace}
\newcommand{\ExamOptwoAcImprovementOverFirstNumCasesLtzeroNod}{\num{0}\xspace}
\newcommand{\ExamOptwoAcImprovementOverFirstNumCasesEqzeroNod}{\num{24}\xspace}

\newcommand{\ExamOptwoAcImprovementOverUnionMeanNod}{\SI{30.0}{\percent}\xspace}
\newcommand{\ExamOptwoAcImprovementOverUnionNumCasesGtzeroNod}{\num{64}\xspace}
\newcommand{\ExamOptwoAcImprovementOverUnionNumCasesLtzeroNod}{\num{0}\xspace}
\newcommand{\ExamOptwoAcImprovementOverUnionNumCasesEqzeroNod}{\num{17}\xspace}

\newcommand{\ExamOptwoAcImprovementOverIndividualMeanNod}{\SI{2.5}{\percent}\xspace}
\newcommand{\ExamOptwoAcImprovementOverIndividualNumCasesGtzeroNod}{\num{13}\xspace}
\newcommand{\ExamOptwoAcImprovementOverIndividualNumCasesLtzeroNod}{\num{6}\xspace}
\newcommand{\ExamOptwoAcImprovementOverIndividualNumCasesEqzeroNod}{\num{62}\xspace}

\newcommand{\ExamBarinelAcImprovementOverFirstMeanNod}{\SI{28.1}{\percent}\xspace}
\newcommand{\ExamBarinelAcImprovementOverFirstNumCasesGtzeroNod}{\num{59}\xspace}
\newcommand{\ExamBarinelAcImprovementOverFirstNumCasesLtzeroNod}{\num{1}\xspace}
\newcommand{\ExamBarinelAcImprovementOverFirstNumCasesEqzeroNod}{\num{21}\xspace}

\newcommand{\ExamBarinelAcImprovementOverUnionMeanNod}{\SI{40.4}{\percent}\xspace}
\newcommand{\ExamBarinelAcImprovementOverUnionNumCasesGtzeroNod}{\num{64}\xspace}
\newcommand{\ExamBarinelAcImprovementOverUnionNumCasesLtzeroNod}{\num{1}\xspace}
\newcommand{\ExamBarinelAcImprovementOverUnionNumCasesEqzeroNod}{\num{16}\xspace}

\newcommand{\ExamBarinelAcImprovementOverIndividualMeanNod}{\SI{39.1}{\percent}\xspace}
\newcommand{\ExamBarinelAcImprovementOverIndividualNumCasesGtzeroNod}{\num{62}\xspace}
\newcommand{\ExamBarinelAcImprovementOverIndividualNumCasesLtzeroNod}{\num{1}\xspace}
\newcommand{\ExamBarinelAcImprovementOverIndividualNumCasesEqzeroNod}{\num{18}\xspace}

\newcommand{\numTests}{\num{101}\xspace}
\newcommand{\numProjects}{\num{48}\xspace}

\newcommand{\numFlakilyExecutedTests}{\num{118}\xspace}
\newcommand{\numFlakilyExecutedOdTests}{\num{29}\xspace}
\newcommand{\numFlakilyExecutedNodTests}{\num{89}\xspace}
\newcommand{\numFlakilyExecutedProjects}{\num{56}\xspace}
\newcommand{\numFlakilyExecutedOdProjects}{\num{5}\xspace}
\newcommand{\numFlakilyExecutedNodProjects}{\num{51}\xspace}
\newcommand{\numFaultNotFoundTests}{\num{17}\xspace}
\newcommand{\numLabeledTestsAllChunks}{\num{227}\xspace}

\newcommand{\numFaultNotFoundHypothesisTestsOrigChunk}{\num{5}\xspace}

\newcommand{\LocationMean}{\num{1.19}\xspace}

\newcommand{\NumProjectsWithMultipleFlakyTest}{\num{12}\xspace}
\newcommand{\NumProjectsWithMultipleFlakyTestAndOneLocation}{\num{8}\xspace}

\newcommand{\LocMedian}{\num{758}\xspace}

\newcommand{\NumFlakyTestsSfflMedian}{\num{2}\xspace}

\newcommand{\NumNonFlakyTestsSfflMedian}{\num{17}\xspace}

\newcommand{\NumNodTests}{\num{81}\xspace}
\newcommand{\NumNodProjects}{\num{45}\xspace}
\newcommand{\NumOdTests}{\num{20}\xspace}

\newcommand{\NumTestsFalselyCategorizedAsOd}{one\xspace}

\newcommand{\InterRaterReliability}{\num{0.66}\xspace}

\newcommand{\ExamTarantulaWcSfflMeanNod}{\num{0.085}\xspace}

\newcommand{\ExamTarantulaWcSfflMedianNod}{\num{0.038}\xspace}

\newcommand{\ExamTarantulaAcSfflMeanNod}{\num{0.064}\xspace}

\newcommand{\ExamTarantulaAcSfflMedianNod}{\num{0.032}\xspace}

\newcommand{\ExamTarantulaBcSfflMeanNod}{\num{0.043}\xspace}

\newcommand{\ExamTarantulaBcSfflMedianNod}{\num{0.018}\xspace}

\newcommand{\ExamOchiaiWcSfflMeanNod}{\num{0.054}\xspace}

\newcommand{\ExamOchiaiWcSfflMedianNod}{\num{0.023}\xspace}

\newcommand{\ExamOchiaiAcSfflMeanNod}{\num{0.035}\xspace}
\newcommand{\ExamOchiaiAcSfflMeanNodHp}{{0.0349}\xspace}

\newcommand{\ExamOchiaiAcSfflMedianNod}{\num{0.014}\xspace}

\newcommand{\ExamOchiaiBcSfflMeanNod}{\num{0.016}\xspace}

\newcommand{\ExamOchiaiBcSfflMedianNod}{\num{0.005}\xspace}

\newcommand{\ExamDstarWcSfflMeanNod}{\num{0.053}\xspace}

\newcommand{\ExamDstarWcSfflMedianNod}{\num{0.023}\xspace}

\newcommand{\ExamDstarAcSfflMeanNod}{\num{0.035}\xspace}
\newcommand{\ExamDstarAcSfflMeanNodHp}{{0.0353}\xspace}
\newcommand{\ExamDstarAcSfflMeanPercentNod}{\SI{3.5}{\percent}\xspace}

\newcommand{\ExamDstarAcSfflMedianNod}{\num{0.014}\xspace}

\newcommand{\ExamDstarBcSfflMeanNod}{\num{0.017}\xspace}

\newcommand{\ExamDstarBcSfflMedianNod}{\num{0.005}\xspace}

\newcommand{\ExamOptwoWcSfflMeanNod}{\num{0.053}\xspace}

\newcommand{\ExamOptwoWcSfflMedianNod}{\num{0.023}\xspace}

\newcommand{\ExamOptwoAcSfflMeanNod}{\num{0.035}\xspace}

\newcommand{\ExamOptwoAcSfflMedianNod}{\num{0.014}\xspace}

\newcommand{\ExamOptwoBcSfflMeanNod}{\num{0.017}\xspace}

\newcommand{\ExamOptwoBcSfflMedianNod}{\num{0.005}\xspace}

\newcommand{\ExamBarinelWcSfflMeanNod}{\num{0.085}\xspace}

\newcommand{\ExamBarinelWcSfflMedianNod}{\num{0.038}\xspace}

\newcommand{\ExamBarinelAcSfflMeanNod}{\num{0.064}\xspace}

\newcommand{\ExamBarinelAcSfflMedianNod}{\num{0.032}\xspace}

\newcommand{\ExamBarinelBcSfflMeanNod}{\num{0.043}\xspace}

\newcommand{\ExamBarinelBcSfflMedianNod}{\num{0.018}\xspace}

\newcommand{\SfflDstarAcRankTenOrLessPercentageNod}{\SI{55.6}{\percent}\xspace}

\newcommand{\ExamTarantulaAcImprovementOverFirstWilcoxonPvalueNodRough}{\num{< 0.001}\xspace}

\newcommand{\ExamTarantulaAcImprovementOverUnionWilcoxonPvalueNodRough}{\num{< 0.001}\xspace}

\newcommand{\ExamTarantulaAcImprovementOverIndividualWilcoxonPvalueNodRough}{\num{< 0.001}\xspace}

\newcommand{\ExamOchiaiAcImprovementOverFirstWilcoxonPvalueNodRough}{\num{< 0.001}\xspace}

\newcommand{\ExamOchiaiAcImprovementOverUnionWilcoxonPvalueNodRough}{\num{< 0.001}\xspace}

\newcommand{\ExamOchiaiAcImprovementOverIndividualWilcoxonPvalueNodRough}{\num{< 0.001}\xspace}

\newcommand{\ExamDstarAcImprovementOverFirstWilcoxonPvalueNodRough}{\num{< 0.001}\xspace}

\newcommand{\ExamDstarAcImprovementOverUnionWilcoxonPvalueNodRough}{\num{< 0.001}\xspace}

\newcommand{\ExamDstarAcImprovementOverIndividualWilcoxonPvalueNodRough}{\num{< 0.001}\xspace}

\newcommand{\ExamOptwoAcImprovementOverFirstWilcoxonPvalueNodRough}{\num{< 0.001}\xspace}

\newcommand{\ExamOptwoAcImprovementOverUnionWilcoxonPvalueNodRough}{\num{< 0.001}\xspace}

\newcommand{\ExamOptwoAcImprovementOverIndividualWilcoxonPvalueNodRough}{\num{< 0.05}\xspace}

\newcommand{\ExamBarinelAcImprovementOverFirstWilcoxonPvalueNodRough}{\num{< 0.001}\xspace}

\newcommand{\ExamBarinelAcImprovementOverUnionWilcoxonPvalueNodRough}{\num{< 0.001}\xspace}

\newcommand{\ExamBarinelAcImprovementOverIndividualWilcoxonPvalueNodRough}{\num{< 0.001}\xspace}

\newcommand{\FlakyTestsMultipleCoveragesPercent}{\SI{80.3}{\percent}\xspace}
\newcommand{\NonFlakyTestsMultipleCoveragesPercent}{\SI{11.9}{\percent}\xspace}

\newcommand{\numFaultNotFoundBySfflTests}{\num{2}\xspace}

\newcommand{\ExamDstarAcSfflNodVsOdMannwhiteneyuEffectsize}{\num{0.90}\xspace}
\newcommand{\ExamDstarAcSfflNetworkVsRandomnessMannwhitneyuPvalue}{\num{0.048}\xspace}
\newcommand{\ExamDstarAcSfflNetworkVsRandomnessMannwhitneyuEffectSize}{\num{0.64}\xspace}

\newcommand{\NumFlakyTestsSfflNetworkMedian}{\num{4}\xspace}

\newcommand{\NumFlakyTestsSfflRandomMedian}{\num{1}\xspace}

\begin{document}

\title{Debugging Flaky Tests using \\Spectrum-based Fault Localization}

\author{%
 \IEEEauthorblockN{Martin Gruber}%
 \IEEEauthorblockA{%
   \textit{BMW Group, University of Passau}\\%
   Munich, Germany\\%
   martin.gr.gruber@bmw.de%
 }%
 \and%
 \IEEEauthorblockN{Gordon Fraser}%
 \IEEEauthorblockA{%
   \textit{University of Passau}\\%
   Passau, Germany\\%
   gordon.fraser@uni-passau.de%
 }
}

\maketitle

\begin{abstract}

Non-deterministically behaving (i.e., \emph{flaky}) tests hamper regression testing as they destroy trust and waste computational and human resources.
Eradicating flakiness in test suites is therefore an important goal, but automated debugging tools are needed to support developers when trying to understand the causes of flakiness.
A popular example for an automated approach to support regular debugging is spectrum-based fault localization (SFL), a technique that identifies software components that are most likely the causes of failures.
While it is possible to also apply SFL for locating likely sources of flakiness in code, unfortunately the flakiness makes SFL both imprecise and non-deterministic.
In this paper we introduce SFFL (Spectrum-based Flaky Fault
Localization), an extension of traditional coverage-based SFL that
exploits our observation that 80\% of flaky tests exhibit varying
coverage behavior between different runs. By distinguishing between
stable and flaky coverage, SFFL is able to locate the sources of
flakiness more precisely and keeps the localization itself
deterministic.
An evaluation on \numTests flaky tests taken from \numProjects open-source Python projects demonstrates that SFFL is effective: Of five prominent SFL formulas, DStar, Ochiai, and Op2 yield the best overall performance.
On average, they are able to narrow down the fault's location to \ExamDstarAcSfflMeanPercentNod of the project's code base, which is \ExamDstarAcImprovementOverFirstMeanNod better than traditional SFL (for DStar).
SFFL's effectiveness, however, depends on the root causes of flakiness: The source of non-order-dependent flaky tests can be located far more precisely than order-dependent faults.

\end{abstract}


\section{Introduction}%
\label{sec:introduction}

Flaky tests are tests that behave non-deterministically, i.e., they can both pass and fail on the same software version.
Test flakiness negatively impacts software development, especially continuous integration~\cite{gruber2022survey,parry2022surveying}:
Before merging their changes, developers want to test them to either find potential bugs early or merge with the confidence that they did not break existing functionality.
The assumption here is that when all tests passed before the change, and one or more tests fail afterwards, these failures are caused by the change and indicate a regression.
For flaky tests, however, this assumption is not true since they may fail intermittently, independently of a change.
Since distinguishing flaky from non-flaky failures is a non-trivial task~\cite{bell2018deflaker,lam2019idflakies,silva2020shake,pinto2020what,alshammari2021flakeflagger,verdecchia2021know,fatima2022flakify,qin2022peeler,parry2022evaluating,li2022evolution}, developers need to investigate flaky failures and inspect the relevant code, mostly only to find that the failure had nothing to do with their changes.

Recent developer surveys confirmed this observation, finding that flakiness wastes developer time and makes it harder to merge pull requests~\cite{gruber2022survey} and that developers who experience flakiness more often become frustrated and are more likely to ignore genuine test failures~\cite{parry2022surveying}.
The latter is particularly problematic since---while most flaky tests are indeed false alarms~\cite{luo2014empirical,eck2019understanding}---test flakiness can also be caused by sporadic bugs in the system under test, which might lead to production failures.
Flaky tests are therefore a pressing issue that needs to be addressed, and ignoring them might lead to severe consequences on developer trust and product quality.

By asking developers what information they require for fixing a flaky test, recent research found that the code elements involved in the flakiness are valuable pieces of information, which, however, are difficult to obtain in complex systems~\cite{eck2019understanding}, and that developers mainly wish for automated debugging tools to better deal with test flakiness~\cite{gruber2022survey}.
We follow this call by adapting spectrum-based fault localization (SFL), a popular and established automated debugging technique, specifically for debugging flaky tests, aiming to identify the exact source code line that causes a test to behave flaky (i.e.,\ the flaky fault).
Unlike traditional SFL, SFFL (Spectrum-based Flaky Fault Localization) takes into account that tests may show different coverage behaviors, which we observe for \FlakyTestsMultipleCoveragesPercent of flaky tests in our evaluation:
For tests that are known to be flaky (e.g.,\ through reruns or other flakiness detection techniques), SFFL only considers a statement as covered, if all test runs covered it.
This follows the intuition that the fault must be located before the failure appears, i.e.,\ before the coverage behavior diverges.
For non-flaky tests on the other hand, it considers a statement as covered if it was involved in any test execution.
The main contributions of this paper are as follows:
\begin{description}
    \item[Approach:] We propose SFFL, a variation of SFL that is specifically designed for debugging flaky tests by considering all observed coverage behaviors of all tests. Our method both increases the effectiveness of fault localization and keeps the fault localization itself mostly deterministic.
    \item[Implementation:] We integrate SFFL in FlaPy~\cite{gruber2021empirical}, a tool for mining flaky tests in a given set of Python projects.
    \item[Dataset:] We provide a dataset of manually labeled fault locations for \numLabeledTestsAllChunks flaky tests.
    \item[Evaluation:] We assess the fault identifying performance of SFFL on our dataset using five established SFL formulas: Tarantula, Ochiai, DStar, Op2, and Barinel.
\end{description}

In our evaluation, the best performing formulas (Ochiai, DStar, and Op2) were able to achieve a mean average-case EXAM score of \ExamDstarAcSfflMeanNod.
That is, a flaky fault could on average be located after inspecting \ExamDstarAcSfflMeanPercentNod of the project's code base.
Compared to the traditional form of SFL that only considers a single coverage behavior for each test, this is a significant improvement of \ExamDstarAcImprovementOverFirstMeanNod, making SFFL practically applicable:
The faults of \SfflDstarAcRankTenOrLessPercentageNod of all flaky tests could be identified by inspecting the top ten ranked statements.

While investigating the fault localization performance on flaky tests of different root causes, we found SFFL to perform notably better on non-order-dependent (NOD) flaky tests than on order-dependent (OD) ones.
This is expected, since OD flakiness can be interpreted as being located either inside other test cases or in the absence of adequate state-resetting code, and both cases are hard to identify for any SFL approach.
There are already specialized approaches for debugging order-dependencies~\cite{gambi2018practical,shi2019ifixflakies,wei2022preempting,wang2022ipflakies}, which however do not apply to NOD flaky tests, such that SFFL provides a great addition and contribution to the overall flakiness debugging toolkit available to developers and researchers.
SFFL can serve as a foundation for further approaches targeting flakiness, such as flakiness detection, program behavior visualization, automated root causing, or automated fixing.
We make all implementations, data, and evaluation scripts available openly~\cite{gruber2023SFFLdataset} and hope that they will foster future research on debugging flaky tests.

\section{Background}%
\label{sec:background}

Fault localization is the process of identifying the component within a system that is responsible for a given (test) failure (i.e., the fault).
Automated fault localization techniques~\cite{wong2016survey} aim to reduce the effort of manual debugging, but also provide a foundation for other automated methods, such as program repair~\cite{gazzola2018automatic}.
Among these, spectrum-based fault localization approaches (SFL) form one of the largest and most popular sub-groups:
Their main idea is to assign each component a suspiciousness score based on the passing and failing executions it was involved in (i.e.,\ the coverage behavior of the tests).
The more failing and the fewer passing executions in which the component participated, the more suspicious it is, i.e., the more likely it is to contain the fault that causes the observed failures.
When sorting the components by their suspiciousness scores, a developer, tester, or automated technique receives a ranked list of components in which the fault is most likely located.
SFL can be applied at various granularities, components can for example be statements, classes, modules, or environment factors (e.g., operating system or no. of CPU cores).

Previous research proposed multiple concrete formulas for calculating the suspiciousness score.
Some popular ones are depicted in \cref{tab:sfl_formulas}.
The term $s$~represents a component (in most cases a \underline{s}tatement), and $S(s)$ denotes its suspiciousness score.
$\mathit{passed}(s)$ and $\mathit{failed}(s)$ are the number of passed/failed executions where $s$ was involved/covered.
$\mathit{totalpassed}$ and $\mathit{totalfailed}$ denote the total number of passed and failed executions.
Note that the value range of Tarantula, Ochiai, and Barinel is $[0,1]$, whereas the value range of DStar is $[0,\infty[$ and the value range of Op2 is $[-\frac{\mathit{totalpassed}}{\mathit{totalpassed} + 1},\mathit{totalfailed}]$.

\setlength{\tabcolsep}{1pt}
\begin{table}
    \centering
    \caption{SFL formulas~\cite{pearson2017evaluating}. For DStar we set the exponent * to 2, as recommended by Wong et al.\cite{wong2014dstar}.}
    \label{tab:sfl_formulas}
\resizebox{\columnwidth}{!}{
\begin{tabular}{ll}
    \toprule
Tarantula~\cite{jones2002visualization}: &
    $S(s)=\frac
    {\mathit{failed}(s)/\mathit{totalfailed}}
    {\mathit{failed}(s)/\mathit{totalfailed} + \mathit{passed}(s)/\mathit{totalpassed}}$
    \vspace{8pt}\\
Ochiai~\cite{abreu2009practical}: &
    $S(s)=\frac
    {\mathit{failed}(s)}
    {\sqrt{\mathit{totalfailed} \cdot (\mathit{failed}(s) + \mathit{passed}(s))}}$
    \vspace{8pt}\\
DStar~\cite{wong2014dstar}: &
    $S(s)=\frac
    {\mathit{failed}(s)^*}
    {\mathit{passed}(s) + (\mathit{totalfailed} - \mathit{failed}(s))}$
    \vspace{8pt}\\
Op2~\cite{naish2011model}: &
    $S(s)=\mathit{failed}(s) - \frac{\mathit{passed}(s)}{\mathit{totalpassed} + 1}$
    \vspace{8pt}\\
Barinel~\cite{abreu2009spectrum}: &
    $S(s)=1 - \frac
    {\mathit{passed}(s)}
    {\mathit{passed}(s) + \mathit{failed}(s)}$
    \\
    \bottomrule
\end{tabular}
}
\end{table}
\setlength{\tabcolsep}{6pt}

Over the past decade, SFL has been applied and evaluated in multiple contexts on both artificial and real faults~\cite{pearson2017evaluating}.
However, the traditional SFL approach assumes the tests to behave deterministically, i.e.,\ to always produce the same test outcome given the same version of the system under test.
This is not the case for flaky tests, which pass and fail intermittently, for example due to networking issues, randomness, or concurrency.
Recent studies have investigated the impact of flaky tests on automated debugging techniques~\cite{cordy2022flakime,qin2021impact,vancsics2020simulating}.
It was found that flakiness corrupts spectrum-based fault localization, resulting in a call for novel fault localization algorithms that take test flakiness into account.
We follow that call by extending SFL for debugging flaky tests, making it more reliable and effective.

To the best of our knowledge, so far there have been two other attempts at applying SFL for debugging flaky tests rather than regular test failures:
Habchi et al.~\cite{habchi2022what} proposed an SFL approach to identify the classes that are responsible for non-deterministic test behavior.
Besides adjusting the SFL formulas for locating the faults of flaky instead of persistently failing tests, they augmented the fault localization with change and code metrics.
FlakyLoc~\cite{moranbarbon2020flakyloc} uses a spectrum-based approach together with combinatorial testing to identify the environmental factors (network bandwidth, browser, screen resolution, \dots) that cause tests in web applications to behave flaky.
However, neither of these approaches aims at identifying the exact statement in the source code that is responsible for flaky test behavior.
Furthermore, they do not utilize the potentially different coverage behaviors of flaky tests, which we identify as a core issue but also an opportunity. %
Our approach (SFFL) addresses both aspects.

The Flakiness Debugger proposed by Ziftci and Cavalcanti~\cite{celalziftci2020de} is another technique aiming at debugging flaky tests.
Like SFFL, it identifies statements related to flaky behavior.
Unlike our approach, however, it is not based on SFL, but on a custom \texttt{DIVERGENCE} algorithm that identifies the first point of divergence in the control flow of each failing run from any of the passing runs.
Therefore, it requires a more complex instrumentation that records the entire dynamic execution trace, for which they use Google-internal technology.
SFFL on the other hand relies only on code coverage information, which can be collected using openly available tools~\cite{hoffmann2011eclemma,pytestCov}.

\newcommand{\myrowcolor}[0]{gray!10}

\begin{table*}
    \centering
    \caption{
        Fault localization matrix of an example test suite illustrating how SFFL works compared to SFL.
        Rows correspond to source code lines.
        $\mathit{totalstable}$ and $\mathit{totalflaky}$ have value 1.
        This code example is similar to the one used by Ziftci et al.~\cite{celalziftci2020de}.
    }
    \label{tab:example}
    \begin{tabular}{|r|l|c|c|c|c|c|c|r|r|r|}
\hline
        &                                        & \multicolumn{2}{c|}{run1} & \multicolumn{2}{c|}{run2} &            &            & SFL       & SFL       & SFFL     \\
*=fault &                                        & t1         & t2           & t1         & t2           & $\cap t1$  & $\cup t2$  & Tarantula & Tarantula & Tarantula \\
        &                                        & pass       & pass         & fail       & pass         & (flaky)    & (stable)   & run1      & run2      & $\cap t1 + \cup t2$  \\\hline
\rowcolor{\myrowcolor}
     1  & |def gen_int():|                       & \checkmark & \checkmark   & \checkmark & \checkmark   & \checkmark & \checkmark & 0.5       & 0.5       & 0.5        \\
     *2 & |    r_int = random.randint(0, 100)|   & \checkmark & \checkmark   & \checkmark & \checkmark   & \checkmark & \checkmark & 0.5       & 0.5       & 0.5        \\
\rowcolor{\myrowcolor}
     *3 & |    if r_int >= 50:|                  & \checkmark & \checkmark   & \checkmark & \checkmark   & \checkmark & \checkmark & 0.5       & 0.5       & 0.5        \\
     4  & |        return "Big number"|          &            &              & \checkmark & \checkmark   &            & \checkmark & 0.0       & 0.5       & 0.0        \\
\rowcolor{\myrowcolor}
     5  & |    else:|                            &            &              &            &              &            & \checkmark & 0.0       & 0.0       & 0.0        \\
     6  & |        return "Small number"|        & \checkmark & \checkmark   &            &              &            & \checkmark & 0.5       & 0.0       & 0.0        \\\cline{3-11}
\rowcolor{\myrowcolor}
     7  & ||\\
     8  & |#flaky test|\\
\rowcolor{\myrowcolor}
     9  & |def t1():|\\
     10 & |    assert "Small" in gen_int()|\\
\rowcolor{\myrowcolor}
     11 & ||\\
     12 & |#stable test|\\
\rowcolor{\myrowcolor}
     13 & |def t2():|\\
     14 & |    assert "number" in gen_int()|\\
\cline{1-2}
    \end{tabular}
\end{table*}

\section{Applying Spectrum-based Fault Localization to Flaky Tests}%
\label{sec:study_setup}

\subsection{Approach}%
\label{sec:Approach}

Unlike deterministically failing tests, flaky tests often have at
least two different coverage behaviors: one for which they pass and
one for which they fail.
For example, consider \cref{tab:example}, which shows a test suite containing two tests, one of them flaky~(\texttt{t1}), the other one stable (\texttt{t2}).
Both tests execute the function \texttt{gen\_int} and make assertions on its return value.
The function itself generates a random integer (line 2) and returns different strings depending on its value (line 3--6).
In this example, the fault that causes \texttt{t1} to behave flaky is
located in lines 2 and 3, because this is where the failure-causing
non-deterministic value is generated (line 2) and where it is
propagated towards the return value and ultimately the assertion (line
3). In contrast, the assertion of \texttt{t2} passes independently of
which string is returned. The test is therefore not flaky.

Arguably, one could also see the fault to be located
in line 10, where the test makes an assertion on a non-deterministic
value.  However, our goal is to find the location where the
failure-causing non-determinism first occurs. %
Whether this non-determinism should be
eradicated, or if the assertion needs to be relaxed, depends on the
specification of the system (oracle problem) and can therefore not be
universally decided.  Knowing the source and location of the
non-determinism causing flaky failures should always be helpful when
debugging flaky tests, regardless of where the fix is applied.

When executing the tests multiple times, we can observe two different coverage behaviors (run1 and run2), which happen to be the same for both tests for this simple example.
In our experiments (\cref{sec:experiment_setup}), we found that
\FlakyTestsMultipleCoveragesPercent of all flaky tests had at least
two different coverage behaviors (i.e., there were statements that
they did not always cover). On the other hand, this was only true for
\NonFlakyTestsMultipleCoveragesPercent of all stable (non-flaky)
tests.
If we perform traditional SFL, for example using Tarantula, lines 1, 2, 3, as well as line 4 (run1) or line 6 (run2) receive the highest suspiciousness scores.
Not only did the fault localization wrongly mark lines 1, 4, and/or 6, the fault localization itself became non-deterministic, making it less usable and potentially subject to the same loss of trust as tests that become flaky.

\setlength{\tabcolsep}{1pt}
\begin{table}[]
    \centering
    \caption{
        SFFL formulas (also used by Habchi et al.~\cite{habchi2022what})
    }
    \label{tab:sfl_formulas_flaky}
\resizebox{\columnwidth}{!}{
\begin{tabular}{ll}
    \toprule
Tarantula~\cite{jones2002visualization}: &
$S(s)=\frac{\mathit{flaky}(s)/\mathit{totalflaky}}{\mathit{flaky}(s)/\mathit{totalflaky} + \mathit{stable}(s)/\mathit{totalstable}}$ \vspace{7pt}\\
Ochiai~\cite{abreu2009practical}: &
    $S(s)=\frac{\mathit{flaky}(s)}{\sqrt{\mathit{totalflaky} \cdot (\mathit{flaky}(s) + \mathit{stable}(s))}}$ \vspace{7pt}\\
DStar~\cite{wong2014dstar}: &
    $S(s)=\frac{\mathit{flaky}(s)^*}{\mathit{stable}(s) + (\mathit{totalflaky} - \mathit{flaky}(s))}$
    \vspace{8pt}\\
Op2~\cite{naish2011model}: &
    $S(s)=\mathit{flaky}(s) - \frac{\mathit{stable}(s)}{\mathit{totalstable} + 1}$
    \vspace{8pt}\\
Barinel~\cite{abreu2009spectrum}: &
    $S(s)=1 - \frac
    {\mathit{stable}(s)}
    {\mathit{stable}(s) + \mathit{flaky}(s)}$
    \\
    \bottomrule
\end{tabular}
}
\end{table}
\setlength{\tabcolsep}{6pt}

The main idea of SFFL is to capitalize on the multi-coverage behavior of flaky tests to narrow down the set of potential fault locations, making it both more effective than SFL and deterministic:
The statement causing the test to behave flaky (the fault) needs to be covered in every test execution, regardless whether the test passes or fails.
A statement that is covered only in passing or only in failing runs can only be the consequence of a flaky fault, but cannot contain the fault itself.
In other words, the faulty non-deterministic value must be created before the coverage behavior diverges.
The fault must therefore be located in the intersection of all coverage behaviors (i.e., covered by both passing and failing executions).
We consider a statement to be covered by a flaky test iff \textit{all} its executions covered it ($\cap t1$).
For stable tests we apply the same logic in opposite direction:
The faulty statement has the potential of making any test flaky that covers it.
A statement that is covered by a stable test is therefore less likely to be faulty.
When a stable test exhibits different coverage behaviors, we take their union, i.e., we consider a statement as covered by a stable test iff \textit{any} of its executions covered it ($\cup t2$).

In detail, SFFL works as follows:
\begin{enumerate}
\item Create the fault localization matrix based on all runs including
  repetitions.
\item Create an error vector in the fault localization matrix based on flakiness (${failed}\rightarrow \mathit{flaky}$, $\mathit{passed}\rightarrow \mathit{stable}$).
\item Add columns for coverage intersection for flaky tests and union for stable ones.
\item Calculate $\mathit{totalflaky}$ and $\mathit{totalstable}$.
\item For each statement $s$ determine $\mathit{stable}(s)$, $\mathit{flaky}(s)$, and calculate its suspiciousness score (\cref{tab:sfl_formulas_flaky}).
\item Rank statements based on suspiciousness scores.
\end{enumerate}

By applying SFFL to our example (last column in \cref{tab:example}),
we can prevent the fault localization from becoming flaky and improve
its effectiveness by eliminating lines 4 and 6 as potential fault
locations.

\subsection{Limitations}%
\label{sec:Limitations}

\subsubsection{Multiple faults}
In our concrete implementation we base SFFL on single fault localization techniques, meaning that our approach assumes one fault per test suite and that its performance might decrease when multiple faults are present.
In our evaluation sample, \NumProjectsWithMultipleFlakyTest out of \numProjects projects contributed more than one flaky test.
From these, \NumProjectsWithMultipleFlakyTestAndOneLocation had the same fault location for all its flaky tests.
This shows that the single fault assumption is not unreasonable, but multiple faults are also present in flaky test suites. In principle, SFFL could also integrate existing work on multiple fault localization (e.g., \cite{abreu2009spectrum}).

\subsubsection{Multiple executions needed}
To apply SFFL, multiple test executions are needed that exhibit both passing and failing test outcomes of the flaky tests.
While this can make a cold start search for flaky faults computationally challenging (as the search for flaky tests itself), it should not be a concern for the practical usage of our approach when debugging an individual flaky test:
The question of where a fault is located usually arises \emph{after} a failure occurred.
When collecting coverage information in every test execution, the failing coverage behavior is therefore already present and all that is needed is the test's coverage behavior in the passing case.
This should be easily retrievable, since (1) flaky tests are mostly rerun anyway~\cite{gruber2022survey,parry2022surveying,micco2016flaky}, and (2) most flaky tests have low failure rates~\cite{gruber2021empirical}, so they are likely to produce a passing execution within a few reruns.
If, however, our test suite contains multiple flaky tests that share the same fault, we need the different coverage behaviors of not one but all flaky tests in the test suite to achieve optimal performance, which might require a substantial number of test executions.
A possible way to achieve this amount of reruns could be by applying DeFlaker~\cite{bell2018deflaker}, a tool that identifies regression test executions that were effectively conducted on the same software version (from the perspective of a certain test case) and can therefore be seen as reruns.

\subsubsection{Coverage needed}
Like the traditional SFL techniques we build on, our approach relies on coverage information, which might be hard to collect in certain scenarios, like the embedded system domain~\cite{elbaum2014techniques}.
Furthermore, the instrumentation needed to retrieve coverage information might also change the behavior of the program and affect the flakiness.
Nevertheless, the concept we present can also be applied to other components (such as lines in log files), or on a more sparse granularity.

\subsubsection{Flakiness in test code}\label{sec:limitations_OD}
SFFL determines the suspiciousness of a statement by the number of flaky and stable tests that execute it.
This makes SFFL less suitable for debugging tests whose flakiness resides in the test code instead of the code under test, which is particularly common for order-dependent flaky tests.

\subsection{Implementation}%
\label{sec:implementation}

To implement SFFL, we extend FlaPy~\cite{gruber2021empirical}, a tool that allows researchers and developers to identify flaky tests in a given set of Python projects by repeatedly executing their test suites.
We chose this option as FlaPy takes care of the isolation of individual test runs, contributing to accurate and realistic results, as well as dependency installation and the parallelization of the experiments, allowing us to conduct our evaluation on a large set of projects.
Furthermore, it provides a convenient interface for collecting and analyzing the test outcomes.

FlaPy relies on pytest~\cite{pytest} to execute the projects' test suites and already uses the pytest-cov plugin~\cite{pytestCov} to generate XML coverage reports, which however do not distinguish between the coverage of individual test cases.
We extend FlaPy by using pytest-cov's \texttt{--cov-context=test} option\cite{pytestCovContext} to record the individual coverage of each test case, which is stored in an sqlite database.
We also extend FlaPy's results-parser to accumulate the coverage information from the different test executions and perform the fault localization.
For order-dependent flaky tests we only consider test executions conducted in random order, and for non-order-dependent flaky tests we only consider those conducted in the same (i.e., default) order.

When implementing the SFL formulas, one has to take division by zero issues into account.
These can occur when the total number of passed executions is zero (Tarantula) or when a statement was covered by all failing and by no passing execution (DStar).
We mitigated division by zero issues by using a save division function that returns zero for 0/0 (an appropriate assumption in our specific case) and infinity for $x$/0 where $x>0$ (negative $x$ values do not occur in the formulas).

\newcommand\notsotiny{\@setfontsize\notsotiny{5.5}{7}}
\lstset{basicstyle=\ttfamily\notsotiny}

\section{Evaluation}%

To assess the efficacy of SFFL, we measure both its effectiveness, and the improvement it yields over traditional SFL.
Namely, we address the following research questions:

\begin{description}

    \item[RQ1 (Effectiveness):] How well can SFFL detect non-order-dependent flaky faults?

    \item[RQ2 (Improvement):] How much better can SFFL locate non-order-dependent flaky faults compared to traditional SFL?

    \item[RQ3 (Root Cause Differences):] How well does SFFL perform on flaky tests of different root causes?

\end{description}

\subsection{Experimental Setup}%
\label{sec:experiment_setup}

To evaluate the fault locating performance of SFFL, we utilize a dataset of flaky tests in Python projects we created in prior work~\cite{gruber2021empirical}, which contains \num{7571} flaky tests from \num{1006} projects.
However, as flakiness is notoriously difficult to reproduce we did not rely on our old findings about which tests are flaky.
Instead, we re-executed the projects' test suites \num{400} times in same order and \num{400} times in random order, which is twice as often as the original study~\cite{gruber2021empirical}.
This large number of reruns provides more confidence in the correct classification of a test as flaky or non flaky, which increases the validity of our evaluation.
We conducted the test executions using our extended version of FlaPy, which collects the individual line coverage for each test case separately (\cref{sec:implementation}).
This process identified \num{5741} flaky tests in \num{476} projects.
\num{4633} of them were identified as order-dependent (OD), \num{874} as non-order-dependent (NOD), and \num{234} as infrastructure flaky.
These numbers are very similar to the ones we reported in our original study, except that now we observed less infrastructure flakiness.
We also found fewer flaky tests overall, however, this can be attributed to the fact that we were unable to execute the test suites of \num{134} of the originally \num{1006} projects.
Despite using the same git revisions of the projects as the original study, this can happen, for example, due to third party libraries becoming incompatible to projects that did not specify the versions of their dependencies (in which case \texttt{pip} installs their latest version).

\subsubsection{Sampling of flaky tests}%
\label{sec:sampling}

\begin{figure}
     \begin{subfigure}[b]{0.58\linewidth}
         \centering
         \includegraphics[width=\linewidth,clip]{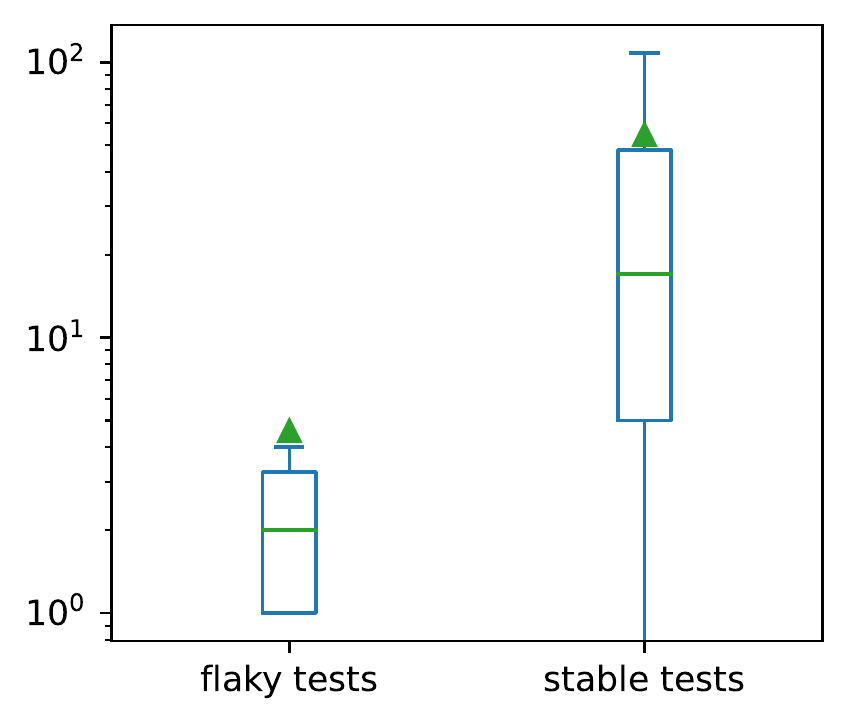}
         \caption{Number of flaky and stable\\tests per project}
         \label{fig:num_tests_per_project}
     \end{subfigure}
     \hfill
     \begin{subfigure}[b]{0.35\linewidth}
         \centering
         \includegraphics[width=\linewidth,clip]{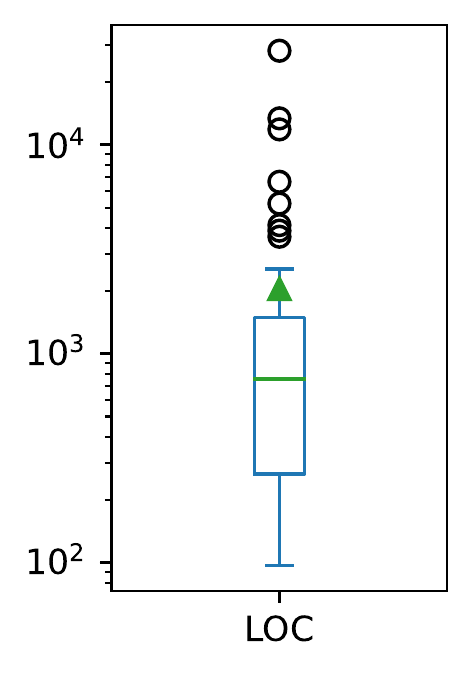}
         \centering
         \caption{Lines of source code per project}
         \label{fig:loc}
     \end{subfigure}
     \caption{Statistics about the \numProjects projects in our dataset. Note the logarithmic y-axes. (triangles=means)}
    \label{fig:project_statistics}
\end{figure}

Our evaluation is based on manually identified fault locations, which are not provided in the original dataset, but labeled by us as part of this study (\cref{sec:ground_truth}).
Since this is an effortful process, we needed to take a sample of the flaky tests we found.
We selected \numFlakilyExecutedNodTests NOD flaky tests from \numFlakilyExecutedNodProjects projects using random stratified sampling, drawing a maximum of ten flaky tests per project to limit the influence of single projects with many flaky tests.
In the same fashion, we selected \numFlakilyExecutedOdTests OD flaky tests from \numFlakilyExecutedOdProjects projects.
We put greater focus on NOD flakiness since SFFL is less qualified for locating OD faults (\cref{sec:limitations_OD}).
In total, we sampled \numFlakilyExecutedTests flaky tests from \numFlakilyExecutedProjects projects.
\Cref{fig:project_statistics} shows statistics about these projects:
The average (median) project contains \NumFlakyTestsSfflMedian flaky and \NumNonFlakyTestsSfflMedian stable tests and has \LocMedian lines of source code.
Note that despite applying random stratified sampling, for the fault localization itself we always consider the entire test suite including all flaky and all stable tests.

\subsubsection{Ground truth / Manual fault localization}%
\label{sec:ground_truth}

To build a ground truth that we can compare our approach against, we performed a manual fault localization:
Using the SFFL results, the error messages, as well as the source code of the respective projects, four software engineering researchers labeled the fault locations of the flaky tests in our sample.
Each flaky test was inspected individually by two researchers.
For cases in which they disagreed, discussions were held until the disagreements were resolved.
According to Fleiss' kappa~\cite{fleiss1971measuring}, we reached an inter-rater reliability of \InterRaterReliability, which is considered \enquote{very good} according to
Regier et al.~\cite{regier2013dsm}.
Multiple fault locations could be assigned to a single flaky test.
On average, the researchers assigned \LocationMean fault locations per flaky test.

For \numFaultNotFoundTests of the \numFlakilyExecutedTests sampled tests, the prerequisites for applying SFFL (\cref{sec:Limitations}), and indeed any form of SFL, were not satisfied, so we had to exclude them.
We discuss these cases in RQ1.
This leaves us with \numTests valid evaluation subjects (flaky tests) from \numProjects projects, of which \NumNodTests are non-order-dependent.%

Apart from the locations of the flaky faults, we also manually determined the root causes of the flaky tests along established categories~\cite{luo2014empirical,eck2019understanding,gruber2021empirical}.
Similar to the study from which we sampled our projects~\cite{gruber2021empirical}%
, we found networking and randomness to be by far the most prevalent causes of NOD flakiness.
The root cause labeling also identified \NumTestsFalselyCategorizedAsOd test that was categorized as order-dependent (only behaved flaky when run in random order, but not in same order), that was actually non-order-dependent flaky, as which we also consider it in the following analyses.
We provide this data openly~\cite{gruber2023SFFLdataset}.

\subsubsection{Metrics}%
\label{sec:metrics}

To determine the performance of spectrum-based flaky fault localization, we compute the suspiciousness rank of each true fault.
For a given statement $s$, its suspiciousness rank $\mathit{rank}(s)$ equals its position in the ranked list of statements, sorted by suspiciousness score in descending order.
In other words, the suspiciousness rank is the number of statements one has to go through until finding the true fault.
The best rank is 1, the worst rank equals the number of lines of code in the program.
Since multiple statements can have the same suspiciousness score, ties can occur.
To accommodate for such cases, we calculate three different ranks: the best-case rank (assuming the statement was listed as the first among its tied neighbors), the worst-case rank (assuming the statement was listed last within the tied group), and the average-case rank as the mean of the best-case and worst-case. %
If multiple true fault locations exist for one test case, the rank of the fault equals the minimum of the ranks of the individual locations, which is the number of statements that need to be processed until the first true fault location is found.

Comparing the suspiciousness ranks between different projects comes with the issue that larger projects (in terms of lines of source code) can skew results, simply because the mere chance of guessing the right statement is smaller for a bigger set of possible statements.
To account for the projects' sizes, we also compute the EXAM score~\cite{wong2008crosstab}, which is the number of source code lines that need to be investigated until a first fault location is found relative to the total number of lines of source code.
In other words, $\mathit{EXAM}(s) = \frac{\mathit{rank}(s)}{N}$, where $N$ is the number of source code lines contained in the respective project.
The value range of the EXAM score is $]0,1]$.
The best EXAM score is 1/$N$, the worst EXAM score is 1.

\subsubsection{RQ1 (Effectiveness)}%

To assess how well SFFL can locate NOD flaky faults, we apply it to the \NumNodTests NOD flaky tests (from \NumNodProjects projects) in our dataset, and compute suspiciousness scores using five established formulas also used in prior empirical studies on spectrum-based fault localization~\cite{pearson2017evaluating}: Tarantula~\cite{jones2002visualization}, Ochiai~\cite{abreu2009practical}, DStar~\cite{wong2014dstar}, Op2~\cite{naish2011model}, and Barinel~\cite{abreu2009spectrum}.
Based on these scores and the manually labeled fault locations, we calculate both the rank and the resulting EXAM scores of each fault.
For counting the projects' lines of source code we used CLOC~\cite{adanial_cloc}.
We compare the outcomes both among each other, determining the best performing formula, and against the performance of traditional SFL on deterministic faults~\cite{pearson2017evaluating}.

\subsubsection{RQ2 (Improvement)}%
\label{sec:setup_rq2}

To measure the effects of SFFL's intersection-union-coverage-processing (\cref{sec:Approach}), we compare it against three baselines:
The \textit{single} baseline performs SFL using only the coverage obtained from the first test execution.
It simulates the circumstances under which Habchi et al.~\cite{habchi2022what} operate, where flaky tests are known, for example through a prediction, however, no historic coverage data is available, so a single execution is used to obtain it.
The \textit{union} baseline performs SFL considering a statement as covered by a test, if it was involved in any of its executions (which SFFL only does for stable tests). It thus considers all available execution data, simulating a naive approach at utilizing multiple coverage behaviors.
The \textit{individual} baseline considers multiple coverage behaviors of the same test case not by aggregating them into one coverage vector, but by treating each test execution like a separate flaky/stable test. $\mathit{totalflaky}$ and $\mathit{totalstable}$ therefore become the number of test executions of flaky and stable tests.
We compare the EXAM scores reached by SFFL against each baseline and use a Wilcoxon signed-rank test~\cite{wilcoxon1945individual}---which is a commonly used, non-parametric paired difference test---to check for statistically relevant differences.

\subsubsection{RQ3 (Root Cause Differences)}%
\label{sec:rq3}

One of the main properties of a flaky test is its root cause, i.e.,\ the broad category of its non-deterministic source, e.g.,\ \textit{networking} or \textit{randomness}.
To investigate if the root cause impacts the effectiveness of SFFL, we compute the EXAM score for each subset of tests sharing the same root cause.
We identify statistically significant differences by using a Mann-Whitney \textit{U} test~\cite{mann1947test} (an unpaired alternative to Wilcoxon) in combination with the Vargha and Delaney $\hat{A}_{12}$ effect size~\cite{vargha2000critique}.
Although we specifically target NOD flaky tests with SFFL, we also include the \NumOdTests OD flaky tests for this research question to validate this design decision.

\subsection{Threats to Validity}%
\label{sec:threats_to_validity}

\subsubsection{External validity}%
\label{sec:external_validity}

The flaky tests we used for our evaluation are taken from an existing dataset of flaky Python tests~\cite{gruber2021empirical}, which was also used by other researchers to evaluate flakiness detection and debugging techniques~\cite{wei2022preempting,wang2022ipflakies,ahmad2022identifying}.
To avoid creating a bias towards projects containing more or less flaky tests, we applied random stratified sampling to derive a tangible subset for our manual fault localization.
Nevertheless, we inherit any potentially existing bias in the original dataset.

\subsubsection{Construct validity}%
\label{sec:construct_validity}

We executed each test 800 times, which is twice as often as the original study that created the dataset.
Nevertheless, some flaky tests have very low failure rates, which might therefore have been misclassified as non-flaky or OD flaky, while being NOD flaky.
Concerning the metrics we use, there are known concerns about whether the EXAM score properly resembles the usefulness of a fault localization technique to a programmer~\cite{parnin2011are}.
However, SFFL is not only targeting manual debugging, but also means to serve as a foundation for other techniques, such as automated~repair.

\subsubsection{Internal validity}%
\label{sec:internal_validity}

Providing the researchers with the SFFL results might influence them during the manual fault localization.
On the other hand, the goal of our evaluation is to verify if SFFL is able to identify locations in the code that are helpful when debugging flakiness and locations where a possible fix \textit{could} be applied.
To judge in which of these locations the fix \textit{should} be applied is only possible with an exact specification, which is not present for most projects.

\subsection{RQ1: Locating NOD Flaky Faults -- Effectiveness}%
\label{sec:rq1}

\begin{table}[]
    \caption{Evaluation results}
    \label{tab:eval_results}
    \centering
    \begin{tabular}{lrrrrrr}
        \toprule
              & \multicolumn{2}{c}{Worst-case}                                      & \multicolumn{2}{c}{Average-case}                            & \multicolumn{2}{c}{Best-case}                          \\
              & Mean                                & Median                        & Mean                                & Median                & Mean                        & Median                   \\
        \midrule
            Tarantula & \ExamTarantulaWcSfflMeanNod & \ExamTarantulaWcSfflMedianNod & \ExamTarantulaAcSfflMeanNod & \ExamTarantulaAcSfflMedianNod & \ExamTarantulaBcSfflMeanNod & \ExamTarantulaBcSfflMedianNod   \\
            Ochiai    & \ExamOchiaiWcSfflMeanNod    & \ExamOchiaiWcSfflMedianNod    & \ExamOchiaiAcSfflMeanNod    & \ExamOchiaiAcSfflMedianNod    & \ExamOchiaiBcSfflMeanNod    & \ExamOchiaiBcSfflMedianNod      \\
            DStar     & \ExamDstarWcSfflMeanNod     & \ExamDstarWcSfflMedianNod     & \ExamDstarAcSfflMeanNod     & \ExamDstarAcSfflMedianNod     & \ExamDstarBcSfflMeanNod     & \ExamDstarBcSfflMedianNod       \\
            Op2       & \ExamOptwoWcSfflMeanNod     & \ExamOptwoWcSfflMedianNod     & \ExamOptwoAcSfflMeanNod     & \ExamOptwoAcSfflMedianNod     & \ExamOptwoBcSfflMeanNod     & \ExamOptwoBcSfflMedianNod       \\
            Barinel   & \ExamBarinelWcSfflMeanNod   & \ExamBarinelWcSfflMedianNod   & \ExamBarinelAcSfflMeanNod   & \ExamBarinelAcSfflMedianNod   & \ExamBarinelBcSfflMeanNod   & \ExamBarinelBcSfflMedianNod     \\
        \bottomrule
    \end{tabular}
\end{table}

\begin{figure}

\centering

\begin{subfigure}[b]{0.49\linewidth}
    \centering
    \includegraphics[width=\linewidth]{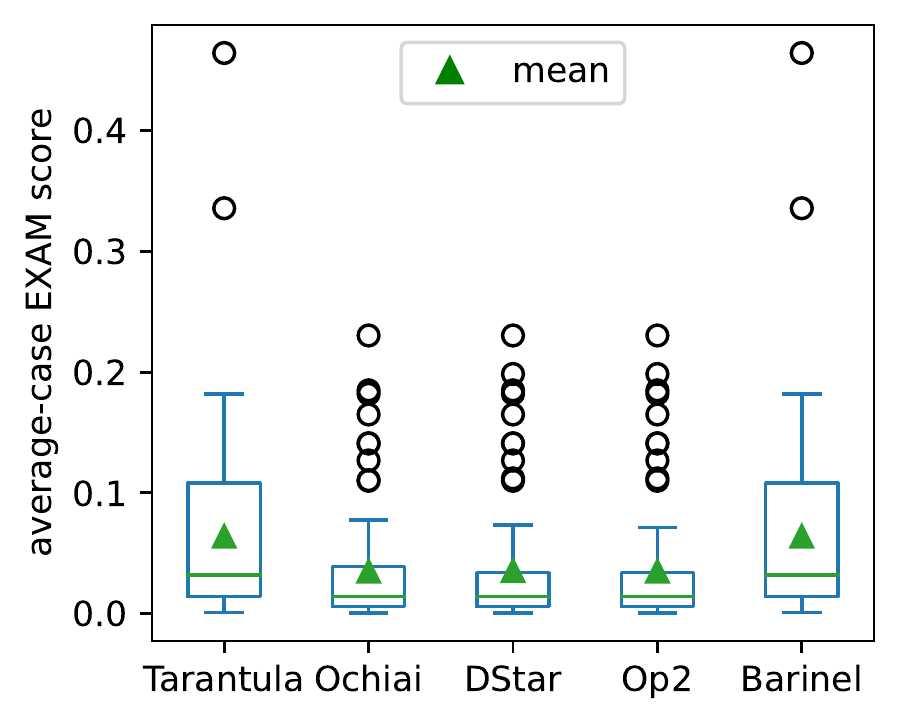}
    \caption{average-case EXAM scores}%
    \label{fig:exam_box}
\end{subfigure}
\hfill
\begin{subfigure}[b]{0.49\linewidth}
    \centering
    \includegraphics[width=\linewidth]{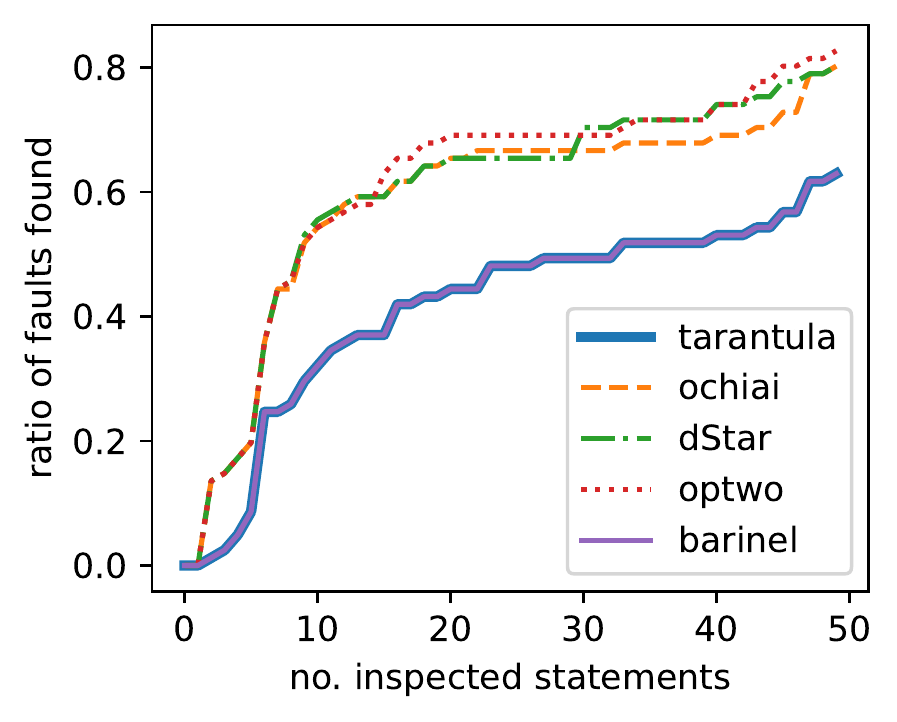}
    \caption{rank progression}%
    \label{fig:rank_progression}
\end{subfigure}

\caption{Effectiveness of SFFL on \NumNodTests NOD flaky tests}

\end{figure}

\cref{tab:eval_results,fig:exam_box} summarize the EXAM scores achieved by SFFL:
Ochiai, DStar, and Op2 yielded the best and also very similar results, whereas Tarantula and Barinel delivered a notably worse performance.
In the following discussions we use DStar as a reference, since it also excelled in other studies~\cite{wong2016survey}.
In terms of absolute EXAM scores, the effectiveness of SFFL is roughly in the same region as the one reached by SFL on deterministic faults~\cite{pearson2017evaluating}, where DStar performed best with a mean EXAM score of 0.0404 (SFFL reached \ExamDstarAcSfflMeanNodHp), followed by Ochiai with 0.0405 (SFFL reached \ExamOchiaiAcSfflMeanNodHp).
The fact that the median scores are always better than the mean scores is caused by poor performing outliers, which are depicted as circles (\cref{fig:exam_box}).
\Cref{fig:rank_progression} shows the progression of faults found after inspecting $x$ statements.
Using DStar, the faults of \SfflDstarAcRankTenOrLessPercentageNod of all flaky tests were listed within the top ten most suspicious statements.
This number underlines the capabilities of SFFL in debugging flaky tests.

The test for which SFFL yielded the best results (average-case DStar EXAM score \num{0.00052}, rank \num{2}) is \texttt{test\_wrap\_strategy} (\cref{fig:example_pygsheet_test}) from project \texttt{pygsheets}\footnote{\url{https://github.com/nithinmurali/pygsheets/tree/01b9548d}}, a Google spreadsheets Python API.
The test case intermittently fails with an HTTP 404 (not found) error.
Given this error message, it is rather obvious that the flakiness is caused by network issues.
However, the test itself does not contain any network interactions, leaving us with the question of where in the system under test the request is made.
SFFL marked three statements (tied) with the highest DStar suspiciousness score for this project: lines 358 and 359 in \texttt{pygsheets/sheet.py} and line 48 in \texttt{pygsheets/spreadsheet.py}.
Indeed the second of these locations is the request causing the HTTP error (\cref{fig:example_pygsheet_fault}).
One explanation for the good performance of SFFL on project \texttt{pygsheets} could be its number of tests:
With 55 flaky and 5 stable tests, it possesses a comparably large test suite and the highest number of flaky test in our evaluation sample.
This gives the fault localization much coverage data to build on, leading to fewer ties in the suspiciousness scores.

\begin{figure}
    \centering

    \begin{subfigure}[b]{\linewidth}
    \centering
\begin{lstlisting}[,firstnumber=768]
def test_wrap_strategy(self):
  cell = self.worksheet.get_values('A1', 'A1', returnas="range")[0][0]
  assert cell.wrap_strategy == "WRAP_STRATEGY_UNSPECIFIED"
  cell.wrap_strategy = "WRAP"
  cell = self.worksheet.get_values('A1', 'A1', returnas="range")[0][0]
  assert cell.wrap_strategy == "WRAP"
\end{lstlisting}
    \caption{
        test\_wrap\_strategy (\texttt{tests/online\_test.py})
    }
    \label{fig:example_pygsheet_test}
    \end{subfigure}

    \begin{subfigure}[b]{\linewidth}

\begin{lstlisting}[firstnumber=350]
def _execute_requests(self, request):
  """Execute a request to the Google Sheets API v4.

  When the API returns a 429 Error will sleep for the specified time and try again.

  :param request:     The request to be made.
  :return:            Response
  """
  try:
\end{lstlisting}
\vspace{-\baselineskip}
\begin{lstlisting}[firstnumber=359,backgroundcolor=\color{yellow}]
    response = request.execute(num_retries=self.retries)
\end{lstlisting}
\vspace{-\baselineskip}
\begin{lstlisting}[firstnumber=360]
  except HttpError as error:
    if error.resp['status'] == '429' and self.check:
      time.sleep(self.seconds_per_quota)  # TODO use asyncio
      response = request.execute(num_retries=self.retries)
    else:
      raise
  return response
\end{lstlisting}

    \centering
    \caption{Flaky fault (\texttt{pygsheets/sheet.py:359})}%
    \label{fig:example_pygsheet_fault}
    \end{subfigure}

    \caption{Project \texttt{pygsheets}}%
    \label{fig:example_pygsheet}
\end{figure}

To illustrate the limitations of SFFL we consider the \numFaultNotFoundTests tests that we excluded from our quantitative evaluation because pytest-cov~\cite{pytestCov} was unable to produce sufficient data, preventing the application of fault localization.
On closer inspection, we found that most of these cases can be attributed to non-deterministic parametrizations that take place before the test function is entered and are therefore not reported as covered by the test.
Specifically, \num{8} of these cases are caused by module-level fixtures and \numFaultNotFoundHypothesisTestsOrigChunk are caused by Hypothesis~\cite{maciver2019hypothesis}, a property-based testing library for Python that dynamically parametrizes a test.
One such case is \texttt{test\_sort\_property} (\cref{fig:example_humansort}) from project \texttt{humansort}\footnote{\url{https://github.com/coreygirard/humansort/tree/270c4ec6}}, a library that offers functionality to sort filenames and other strings in a human-readable way.
The test case intermittently fails due to timeouts caused by Hypothesis.
We marked line 122 as the fault location, since this is where the parametrization takes place.
However, since this happens \textit{before} the test function is entered, pytest-cov does not consider it to be covered by the test itself, which is why the statement is not part of the suspiciousness matrix.
Since these issues stem from the coverage collection, this is not a limitation of SFFL specifically, but a general concern for any SFL approach.

\begin{figure}
    \begin{lstlisting}[firstnumber=6]
from hypothesis import given
from hypothesis.strategies import from_regex, integers, lists, tuples
# ... omitted ... `\Suppressnumber`
`\Reactivatenumber{115}`
# Create filename "templates", ie before digits
strat_strings = from_regex(r"\A[^0-9]*\Z")
strat_mod = tuples(integers(), lists(integers(min_value=0), max_size=10))

strat = strat_strings, lists(strat_mod, max_size=5)


\end{lstlisting}
\vspace{-\baselineskip}
\begin{lstlisting}[firstnumber=122,backgroundcolor=\color{yellow}]
@given(lists(tuples(*strat)))
\end{lstlisting}
\vspace{-\baselineskip}
\begin{lstlisting}[firstnumber=123]
def test_sort_property(e):
    strings = gen_test_data(e)
    backup = strings[:]
    random.shuffle(strings)
    assert backup == sort(strings)
\end{lstlisting}
    \caption{Flaky test in project \texttt{humansort} (\texttt{tests/test\_sort.py})}%
    \label{fig:example_humansort}
\end{figure}

\summary{RQ1: Effectiveness}{
    Using the DStar formula, SFFL was able to achieve a mean average-case EXAM score of \ExamDstarAcSfflMeanNod (median \ExamDstarAcSfflMedianNod) and locate the faults of \SfflDstarAcRankTenOrLessPercentageNod of all flaky tests within ten statements.
}

\subsection{RQ2: Locating NOD Flaky Faults -- Improvement}%
\label{sec:rq2_comparison_to_un_tuned_fault_localization}

\begin{table*}
    \centering
    \caption{
        SFFL vs.\ baselines by average-case EXAM scores.
    }
    \label{tab:improvement_over_baselines}

    \begin{tabular}{lrcrlcrlcrl}
        \toprule
        {} & {\underline{\smash{SFFL}}} &
        \multicolumn{3}{|c}{\underline{\smash{baseline = \textit{single}}}} &
        \multicolumn{3}{|c}{\underline{\smash{baseline = \textit{union}}}} &
        \multicolumn{3}{|c}{\underline{\smash{baseline = \textit{individual}}}}
        \\
        \multilinecellL{SFL\\formula} &
        \multilinecellR{mean\\av.-case\\EXAM\\score} &
        \multicolumn{1}{|c}{\multilinecellC{SFFL better,\\unchanged,\\SFFL worse,\\not found}} &
            \multilinecellR{mean\\impr.} &
            \multilinecellL{Wilcoxon\\p-value} &
        \multicolumn{1}{|c}{\multilinecellC{SFFL better,\\unchanged,\\SFFL worse,\\not found}} &
            \multilinecellR{mean\\impr.} &
            \multilinecellL{Wilcoxon\\p-value} &
        \multicolumn{1}{|c}{\multilinecellC{SFFL better,\\unchanged,\\SFFL worse,\\not found}} &
            \multilinecellR{mean\\impr.} &
            \multilinecellL{Wilcoxon\\p-value}
        \\
        \midrule
        Tarantula &
            \ExamTarantulaAcSfflMeanNod
            &
            \ExamTarantulaAcImprovementOverFirstNumCasesGtzeroNod,
            \ExamTarantulaAcImprovementOverFirstNumCasesEqzeroNod,
            \ExamTarantulaAcImprovementOverFirstNumCasesLtzeroNod,
            \numFaultNotFoundBySfflTests &
            \ExamTarantulaAcImprovementOverFirstMeanNod &
            \ExamTarantulaAcImprovementOverFirstWilcoxonPvalueNodRough
            &
            \ExamTarantulaAcImprovementOverUnionNumCasesGtzeroNod,
            \ExamTarantulaAcImprovementOverUnionNumCasesEqzeroNod,
            \ExamTarantulaAcImprovementOverUnionNumCasesLtzeroNod,
            \numFaultNotFoundBySfflTests &
            \ExamTarantulaAcImprovementOverUnionMeanNod &
            \ExamTarantulaAcImprovementOverUnionWilcoxonPvalueNodRough
            &
            \ExamTarantulaAcImprovementOverIndividualNumCasesGtzeroNod,
            \ExamTarantulaAcImprovementOverIndividualNumCasesEqzeroNod,
            \ExamTarantulaAcImprovementOverIndividualNumCasesLtzeroNod,
            \numFaultNotFoundBySfflTests &
            \ExamTarantulaAcImprovementOverIndividualMeanNod &
            \ExamTarantulaAcImprovementOverIndividualWilcoxonPvalueNodRough
            \\
        Ochiai &
            \ExamOchiaiAcSfflMeanNod
            &
            \ExamOchiaiAcImprovementOverFirstNumCasesGtzeroNod,
            \ExamOchiaiAcImprovementOverFirstNumCasesEqzeroNod,
            \ExamOchiaiAcImprovementOverFirstNumCasesLtzeroNod,
            \numFaultNotFoundBySfflTests &
            \ExamOchiaiAcImprovementOverFirstMeanNod &
            \ExamOchiaiAcImprovementOverFirstWilcoxonPvalueNodRough
            &
            \ExamOchiaiAcImprovementOverUnionNumCasesGtzeroNod,
            \ExamOchiaiAcImprovementOverUnionNumCasesEqzeroNod,
            \ExamOchiaiAcImprovementOverUnionNumCasesLtzeroNod,
            \numFaultNotFoundBySfflTests &
            \ExamOchiaiAcImprovementOverUnionMeanNod &
            \ExamOchiaiAcImprovementOverUnionWilcoxonPvalueNodRough
            &
            \ExamOchiaiAcImprovementOverIndividualNumCasesGtzeroNod,
            \ExamOchiaiAcImprovementOverIndividualNumCasesEqzeroNod,
            \ExamOchiaiAcImprovementOverIndividualNumCasesLtzeroNod,
            \numFaultNotFoundBySfflTests &
            \ExamOchiaiAcImprovementOverIndividualMeanNod &
            \ExamOchiaiAcImprovementOverIndividualWilcoxonPvalueNodRough
            \\
        DStar &
            \ExamDstarAcSfflMeanNod
            &
            \ExamDstarAcImprovementOverFirstNumCasesGtzeroNod,
            \ExamDstarAcImprovementOverFirstNumCasesEqzeroNod,
            \ExamDstarAcImprovementOverFirstNumCasesLtzeroNod,
            \numFaultNotFoundBySfflTests &
            \ExamDstarAcImprovementOverFirstMeanNod &
            \ExamDstarAcImprovementOverFirstWilcoxonPvalueNodRough
            &
            \ExamDstarAcImprovementOverUnionNumCasesGtzeroNod,
            \ExamDstarAcImprovementOverUnionNumCasesEqzeroNod,
            \ExamDstarAcImprovementOverUnionNumCasesLtzeroNod,
            \numFaultNotFoundBySfflTests &
            \ExamDstarAcImprovementOverUnionMeanNod &
            \ExamDstarAcImprovementOverUnionWilcoxonPvalueNodRough
            &
            \ExamDstarAcImprovementOverIndividualNumCasesGtzeroNod,
            \ExamDstarAcImprovementOverIndividualNumCasesEqzeroNod,
            \ExamDstarAcImprovementOverIndividualNumCasesLtzeroNod,
            \numFaultNotFoundBySfflTests &
            \ExamDstarAcImprovementOverIndividualMeanNod &
            \ExamDstarAcImprovementOverIndividualWilcoxonPvalueNodRough
            \\
        Op2 &
            \ExamOptwoAcSfflMeanNod
            &
            \ExamOptwoAcImprovementOverFirstNumCasesGtzeroNod,
            \ExamOptwoAcImprovementOverFirstNumCasesEqzeroNod,
            \ExamOptwoAcImprovementOverFirstNumCasesLtzeroNod,
            \numFaultNotFoundBySfflTests &
            \ExamOptwoAcImprovementOverFirstMeanNod &
            \ExamOptwoAcImprovementOverFirstWilcoxonPvalueNodRough
            &
            \ExamOptwoAcImprovementOverUnionNumCasesGtzeroNod,
            \ExamOptwoAcImprovementOverUnionNumCasesEqzeroNod,
            \ExamOptwoAcImprovementOverUnionNumCasesLtzeroNod,
            \numFaultNotFoundBySfflTests &
            \ExamOptwoAcImprovementOverUnionMeanNod &
            \ExamOptwoAcImprovementOverUnionWilcoxonPvalueNodRough
            &
            \ExamOptwoAcImprovementOverIndividualNumCasesGtzeroNod,
            \ExamOptwoAcImprovementOverIndividualNumCasesEqzeroNod,
            \ExamOptwoAcImprovementOverIndividualNumCasesLtzeroNod,
            \numFaultNotFoundBySfflTests &
            \ExamOptwoAcImprovementOverIndividualMeanNod &
            \ExamOptwoAcImprovementOverIndividualWilcoxonPvalueNodRough
            \\
        Barinel &
            \ExamBarinelAcSfflMeanNod
            &
            \ExamBarinelAcImprovementOverFirstNumCasesGtzeroNod,
            \ExamBarinelAcImprovementOverFirstNumCasesEqzeroNod,
            \ExamBarinelAcImprovementOverFirstNumCasesLtzeroNod,
            \numFaultNotFoundBySfflTests &
            \ExamBarinelAcImprovementOverFirstMeanNod &
            \ExamBarinelAcImprovementOverFirstWilcoxonPvalueNodRough
            &
            \ExamBarinelAcImprovementOverUnionNumCasesGtzeroNod,
            \ExamBarinelAcImprovementOverUnionNumCasesEqzeroNod,
            \ExamBarinelAcImprovementOverUnionNumCasesLtzeroNod,
            \numFaultNotFoundBySfflTests &
            \ExamBarinelAcImprovementOverUnionMeanNod &
            \ExamBarinelAcImprovementOverUnionWilcoxonPvalueNodRough
            &
            \ExamBarinelAcImprovementOverIndividualNumCasesGtzeroNod,
            \ExamBarinelAcImprovementOverIndividualNumCasesEqzeroNod,
            \ExamBarinelAcImprovementOverIndividualNumCasesLtzeroNod,
            \numFaultNotFoundBySfflTests &
            \ExamBarinelAcImprovementOverIndividualMeanNod &
            \ExamBarinelAcImprovementOverIndividualWilcoxonPvalueNodRough
            \\
        \bottomrule
    \end{tabular}

\end{table*}

\begin{figure}
    \centering
    \includegraphics[width=\linewidth]{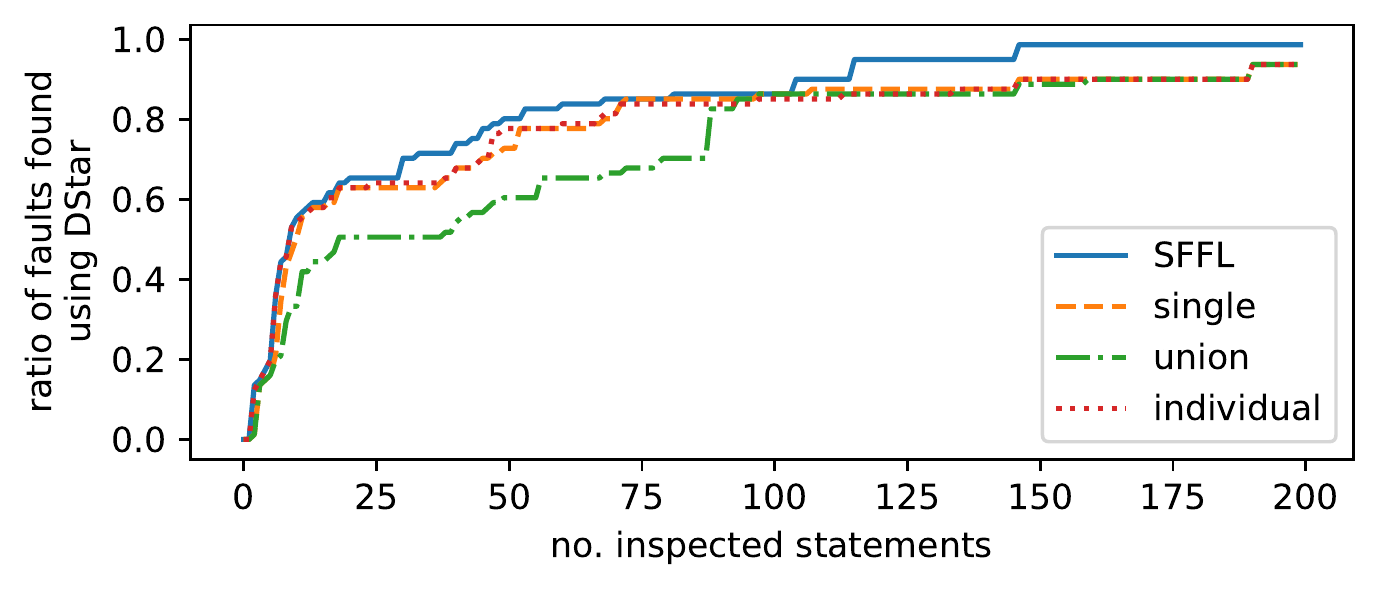}
    \caption{Rank progression of DStar compared to baselines}%
    \label{fig:rank_progression_improvement}
\end{figure}

\cref{tab:improvement_over_baselines} depicts the improvement yielded by SFFL over our three baselines (\cref{sec:setup_rq2}).
To quantify the overall improvement, we use the mean change in the EXAM score.
We chose the mean over the median since latter is often zero for the \textit{individual} baseline, despite SFFL yielding significant improvements.
In fact, across all formulas and baselines, SFFL significantly improved the fault locating performance (p-value $< 0.05$).
The superiority of SFFL can also be observed in \cref{fig:rank_progression_improvement}, which shows the rank progression for all baselines:
SFFL always performs at least as strong as the best baseline.
Notably, the \textit{union} baseline performed worse than the \textit{single} baseline, which a Wilcoxon signed-rank test confirmed (p $< 0.001$).
This shows that a naive usage of multiple coverage behaviors does not improve but may even diminish the effectiveness of SFL on flaky tests.

\begin{figure}
    \centering

    \begin{subfigure}[b]{\linewidth}
    \centering
\begin{lstlisting}[firstnumber=39]
@pytest.fixture()
def session():
    """Create a basic aiobfd session"""
    return aiobfd.session.Session('127.0.0.1', '127.0.0.1')`\Suppressnumber`
    # ... omitted ...
`\Reactivatenumber{217}`
def test_sess_rx_interval_get(session):
    """Attempt to get the Required Min Rx Interval"""
    assert session.required_min_rx_interval == 1000000
\end{lstlisting}
    \caption{test\_sess\_rx\_interval\_get (\texttt{tests/test\_session.py})}
    \label{fig:example_aiobf_test}
    \end{subfigure}

    \begin{subfigure}[b]{\linewidth}
\begin{lstlisting}[firstnumber=49]
class Session:
  """BFD session with a remote"""

  def __init__(self, local, remote, family=socket.AF_UNSPEC, passive=False,
\end{lstlisting}
\vspace{-\baselineskip}
\begin{lstlisting}[firstnumber=53]
    tx_interval=1000000, rx_interval=1000000, detect_mult=3):`\Suppressnumber`
    # ... omitted ...
    `\Reactivatenumber{91}`
    # Create the local client and run it once to grab a port
    log.debug('Setting up UDP client for %s:%s.', remote, CONTROL_PORT)
\end{lstlisting}
\vspace{-\baselineskip}
\begin{lstlisting}[firstnumber=93,backgroundcolor=\color{yellow}]
    src_port = random.randint(SOURCE_PORT_MIN, SOURCE_PORT_MAX)
\end{lstlisting}
\vspace{-\baselineskip}
\begin{lstlisting}[firstnumber=94]
    fam, _, _, _, addr = socket.getaddrinfo(self.local, src_port)[0]`\Suppressnumber`
    # ... omitted ... `\Reactivatenumber{102}`
\end{lstlisting}
\vspace{-\baselineskip}
\begin{lstlisting}[firstnumber=102]
    sock.bind(addr)
\end{lstlisting}
    \centering
    \caption{Flaky fault (\texttt{aiobfd/session.py:93)}}
    \label{fig:example_aiobf_fault}
    \end{subfigure}

    \caption{Project \texttt{aiobfd}}%
    \label{fig:name}
\end{figure}

\begin{table}

    \caption{Suspiciousness matrices for project \texttt{aiobfd}}

\begin{subtable}{\linewidth}
    \caption{using traditional SFL}
    \label{tab:sfl_matrix_aiobfd}
    \centering
    \resizebox{\linewidth}{!}{
    \begin{tabular}{llll}
        \toprule
        rank                   & file (* = contains fault)         & line         & \multilinecell{DStar susp. score}  \\
        \midrule
        1,2                    & \phantom{*}\texttt{tests/test\_session.py} & 219,353      & 1.0                              \\
        3-47                   & \phantom{*}\texttt{aiobfd/session.py}      & 148$\sim$461 & 0.0727                           \\
        \multirow{2}{*}{48-95} & *\texttt{aiobfd/session.py}                & 55-112,123   & \multirow{2}{*}{0.0714}          \\
                               & \phantom{*}\texttt{tests/test\_session.py} & 42           &                                  \\
        96,97                  & \phantom{*}\texttt{aiobfd/transport.py}    & 12,16        & 0.0702                           \\
        \bottomrule
    \end{tabular}
    }
\end{subtable}%

\vspace{12pt}

\begin{subtable}{\linewidth}
    \caption{using SFFL}
    \label{tab:sffl_matrix_aiobfd}
    \centering
    \resizebox{\linewidth}{!}{
    \begin{tabular}{llllr}
        \toprule
        rank                                & file (* = contains fault)         & line                & \multilinecell{DStar susp. score}  \\
        \midrule
        \multirow{2}{*}{1-39\phantom{0}}    & *\texttt{aiobfd/session.py}                & 55-102\phantom{123} & \multirow{2}{*}{0.0714}          \\
                                            & \phantom{*}\texttt{tests/test\_session.py} & 42                  & \\
        \bottomrule
    \end{tabular}
    }
\end{subtable}

\end{table}

The fault on which SFFL exhibits the largest improvement (\SI{72}{\percent} for DStar over the \textit{single} baseline) was \texttt{test\_sess\_rx\_interval\_get}~(\cref{fig:example_aiobf_test}) from project \texttt{aiobfd}\footnote{\url{https://github.com/netedgeplus/aiobfd/tree/09012247}}, a library implementing the BFD network protocol.
The test case itself is rather short and its flakiness comes from its fixture%
\footnote{native pytest fixtures are not affected by the coverage-collection issue that \texttt{humansort} faced (\cref{fig:example_humansort}). They are reported as covered by the respective test.}%
, which creates a Session object using a random integer value for the port number (\cref{fig:example_aiobf_fault}, line 93).
The test case fails intermittently in case the port is already in use, leading to an \texttt{OSError: [Errno 98] Address already in use} in line \num{102}.
In terms of root cause, this test case shows a mixture of local networking and randomness and we consider line 93 of the session module (\cref{fig:example_aiobf_fault}) as the faulty statement.

\Cref{tab:sfl_matrix_aiobfd} depicts the corresponding ranked fault localization matrix using the \textit{single} method.
The two statements receiving the highest suspiciousness scores are the test code itself (line 219), as well as the code of the one other flaky test in the project (line 353).
The following block (rank 3--47) contains multiple methods of the Session class.
Then follows the constructor of the Session class (lines 55--112)---which contains the fault---as well as another method (line 123), and the session fixture (line~42).
The first block (rank 3--47) received a slightly higher suspiciousness score, because it was covered by fewer non-flaky tests than the constructor.
The faulty statement therefore receives an average-case suspiciousness rank of \num{71.5} (best-case \num{48}, worst-case \num{95}).
\Cref{tab:sffl_matrix_aiobfd} shows the ranked fault localization matrix using SFFL.
By only considering statements that were covered by both passing and failing executions of flaky tests, the approach was able to cut down the number of possibly faulty statements from \num{97} to \num{39}, crossing out the test code itself, the part of the constructor that is executed after the failure appeared (line 102), as well as other methods of the Session class.
This project is a great example for how combining the coverage behavior of multiple test executions can increase the focus on actual faults.

In %
two special cases, SFFL was not able to locate the fault at all, although the baselines could, which was caused by its coverage intersection processing.
In particular, the projects of these two cases are called \texttt{mpesa-py}\footnote{\url{https://github.com/Arlus/mpesa-py/tree/2844ac9c}} and \texttt{pympesa}, and both flaky tests are named \texttt{test\_get\_balance}.
After manual inspection we found that they are probably undeclared forks.
According to their README, they implement a wrapper providing convenient access to the API of MPESA, a mobile phone-based money transfer service.

The test creates a Balance object
(\cref{fig:example_mpesa_py}) and calls the \texttt{get\_balance} method, which
issues a remote API request (line 50).  Intermittent failures occur in
case the web service is not available.  However, network issues can
also occur earlier during the authentication (line 9), which is in
fact the cause for six other flaky tests in the projects.  Since
executions that experienced authentication failures did not reach line
50, SFFL did not consider the line to be consistently covered by the
flaky test.  As no other test covered it either, line 50 was excluded
from the suspiciousness matrix and the fault was not found by SFFL.
Consequently, this is a multi-fault localization problem, which is a
well known limitation of SFL~\cite{abreu2009spectrum}, but the effect
was amplified by SFFL's coverage processing.

\begin{figure}
    \centering
\begin{lstlisting}[firstnumber=1,breakindent=25pt]
import requests
from .auth import MpesaBase


class Balance(MpesaBase):
  def __init__(self, env="sandbox", app_key=None, app_secret=None, sandbox_url="https://sandbox.safaricom.co.ke",
\end{lstlisting}
\vspace{-\baselineskip}
\begin{lstlisting}[firstnumber=7,breakindent=25pt]
               live_url="https://safaricom.co.ke"):
    MpesaBase.__init__(self, env, app_key, app_secret, sandbox_url, live_url)
\end{lstlisting}
\vspace{-\baselineskip}
\begin{lstlisting}[firstnumber=9,breakindent=25pt,backgroundcolor=\color{yellow}]
    self.authentication_token = self.authenticate()
\end{lstlisting}
\vspace{-\baselineskip}
\begin{lstlisting}[firstnumber=10,breakindent=25pt]

  def get_balance(self, initiator=None, security_credential=None, command_id=None, party_a=None, identifier_type=None,
\end{lstlisting}
\vspace{-\baselineskip}
\begin{lstlisting}[firstnumber=12,breakindent=25pt]
    remarks=None, queue_timeout_url=None,result_url=None):`\Suppressnumber`
    # ... omitted ...`\Reactivatenumber{50}`
\end{lstlisting}
\vspace{-\baselineskip}
\begin{lstlisting}[firstnumber=50,breakindent=25pt,backgroundcolor=\color{yellow}]
    r = requests.post(saf_url, headers=headers, json=payload)
\end{lstlisting}
\vspace{-\baselineskip}
\begin{lstlisting}[firstnumber=51,breakindent=25pt]
    return r.json()
\end{lstlisting}
\caption{Project \texttt{mpesa-py} (\texttt{mpesa/api/balance.py})}
    \label{fig:example_mpesa_py}
\end{figure}

\summary{RQ2: Improvement}{%
    SFFL significantly improves fault localization on flaky tests.
    For DStar, the EXAM score improved in \ExamDstarAcImprovementOverFirstNumCasesGtzeroNod out of \NumNodTests cases and by a mean of \ExamDstarAcImprovementOverFirstMeanNod compared using a single test execution.
    When facing multiple faults, SFFL can however fail to locate all of them.
}

\subsection{RQ3: Root Cause Differences}%
\label{sec:rq3_sffl_on_order_dependencies}

\begin{figure}
    \centering
    \includegraphics[width=\linewidth]{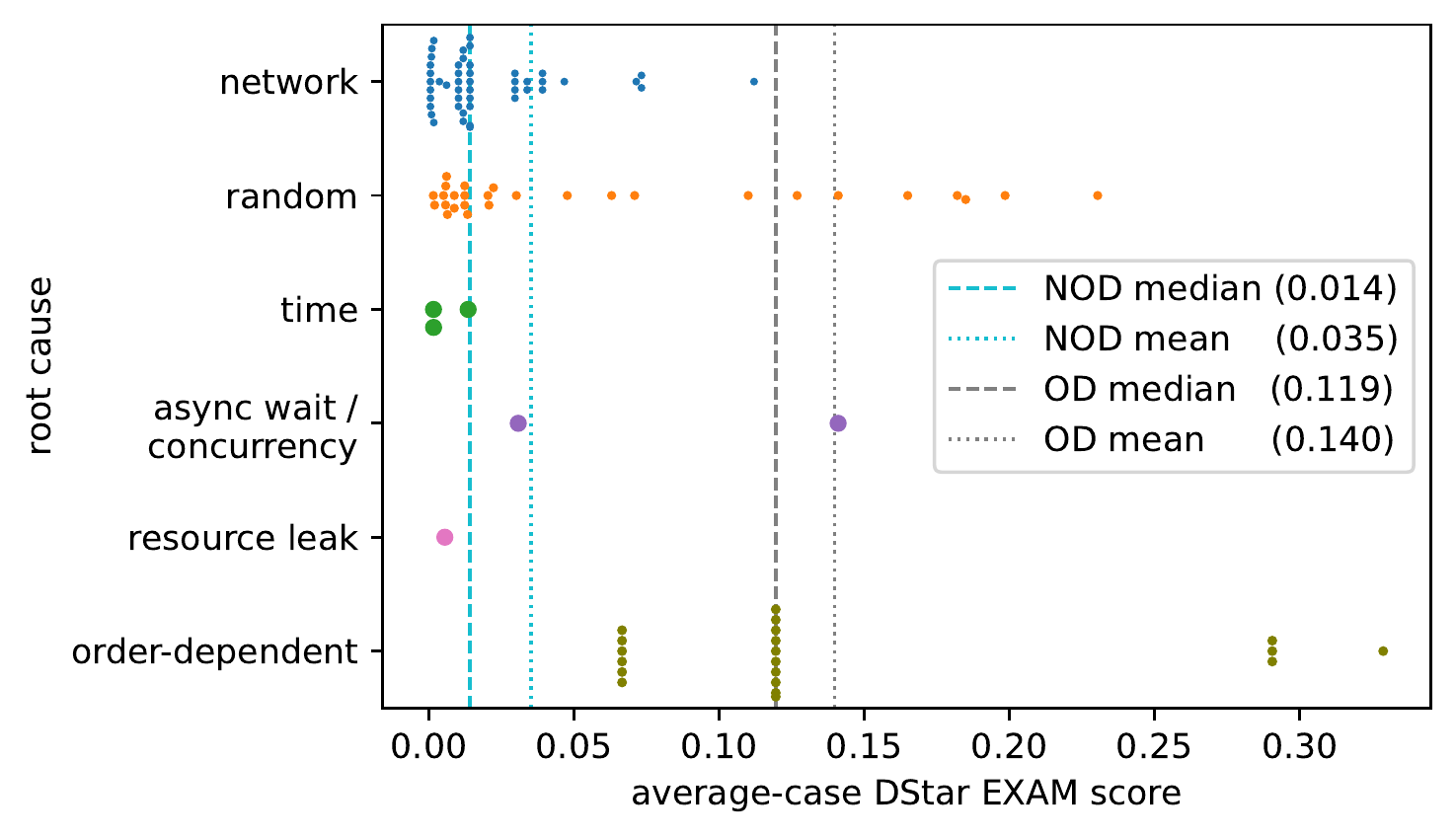}
    \caption{
        Swarm plot of EXAM scores grouped by root cause
    }%
    \label{fig:root_cause_swarm}
\end{figure}

\Cref{tab:root_causes,fig:root_cause_swarm} show the distribution of tests across root cause categories as well as the performance reached by SFFL on the respective subsets.
While we found only small differences among NOD root causes, we see a strong gap between NOD and OD flaky tests:
Latter differ significantly (p-value of Mann–Whitney \textit{U} test $< 0.001$) with a large effect size ($\hat{A}_{12}$ = \ExamDstarAcSfflNodVsOdMannwhiteneyuEffectsize). %
This confirms our intuition that SFFL is not well suited for OD tests, since it relies on many failing/flaky tests covering faulty statements, which is usually not the case for faults that are located inside a test case.
For the same reason, Habchi et al.~\cite{habchi2022what} excluded OD flaky tests from their evaluation of SFL on flaky tests.
While order-dependency is a frequent cause of flakiness, we regard SFFL's confined ability to debug OD flaky tests as a negligible limitation, since specialized approaches for debugging order-dependencies exist%
~\cite{gambi2018practical,shi2019ifixflakies,wei2022preempting,wang2022ipflakies}.
NOD flaky tests, on the other, have not seen the same attention in terms of debugging approaches, which SFFL addresses.

For NOD flaky tests, we find significant differences only between network- and randomness-related flaky tests
(p = \ExamDstarAcSfflNetworkVsRandomnessMannwhitneyuPvalue). The effect size is small ($\hat{A}_{12}$ = \ExamDstarAcSfflNetworkVsRandomnessMannwhitneyuEffectSize), and the difference is most likely caused by projects of the former simply containing more flaky tests on average (median~\NumFlakyTestsSfflNetworkMedian) than projects with randomness-related flakiness (median~\NumFlakyTestsSfflRandomMedian), which helps SFFL to locate faults.
Therefore, developers should be able to expect a similar performance from SFFL on NOD flaky tests, regardless of their exact root cause, which does also not need to be known to apply SFFL.

\summary{RQ3: Root Cause Differences}{
    SFFL performs expectedly worse on order-dependent flaky tests compared to non-order-dependent ones.
    Among different non-order-dependent root causes we found no substantial differences.
}

\begin{table}
    \centering
    \caption{Performance differences between root causes}
    \label{tab:root_causes}
    \resizebox{\linewidth}{!}{
        \begin{tabular}{lrrrr}
\toprule
              root cause & no. tests & \multicolumn{3}{l}{\multilinecell{\underline{\smash{average-case\ DStar EXAM score}}}} \\
                         &        {} &                                                             median &  mean &   std \\
\midrule
                 network &        47 &                                              0.014 & 0.020 & 0.023 \\
                  random &        28 &                                              0.021 & 0.061 & 0.073 \\
         order-dependent &        20 &                                              0.119 & 0.140 & 0.086 \\
                    time &         3 &                                              0.002 & 0.006 & 0.007 \\
async wait / concurrency &         2 &                                              0.086 & 0.086 & 0.078 \\
           resource leak &         1 &                                              0.005 & 0.005 &   - \\
\bottomrule
\end{tabular}

    }
\end{table}

\section{related work}%
\label{sec:related_work}

\subsection{The Impact of Flakiness on SFL}%
\label{sec:the_impact_of_flakiness_on_sfl}

Several prior studies investigated the effects of flaky tests on fault localization:
Cordy et al.~\cite{cordy2022flakime} investigated the impact of flakiness on automated program repair (of which fault localization is the first step) using Defects4J~\cite{just2014defects4j} %
and found that the fault localization step is particularly sensitive to test flakiness:
Even the smallest degree of flakiness was sufficient to diminish both precision and recall of threshold-based suspicious statement selection by more than \SI{80}{\percent}.
This can have disastrous effects on patch generation, which was independently confirmed by another study on automatic repair~\cite{martinez2017automatic}.
Qin et al.~\cite{qin2021impact} also studied the impact of flaky tests on automated program repair based on
tests that have been annotated as flaky in the Defects4J defect benchmark~\cite{just2014defects4j}, and also found
flakiness to significantly decrease fault localization efficacy, with localization results decreasing for \num{90} out of \num{387} bugs and none improving after including flaky tests.
Vancsics et al.\cite{vancsics2020simulating} simulated the effect of flakiness on fault localization using \num{28} bugs from the Defects4J dataset~\cite{just2014defects4j}.
They found that after adding flakiness (without changing coverage), most bugs received a suspiciousness rank that was within $\pm25\%$ of the original.
All of these studies motivate the need to investigate SFL in the face of flakiness, which we do in this paper by proposing SFFL.
Since SFFL produces consistent results despite varying test outcomes, it can absorb the effect of flakiness and form a solid foundation for automated program repair and other techniques building on fault localization.

\subsection{Flakiness Debugging}%
\label{sec:coverage_based_flakiness_debugging}

Habchi et al.~\cite{habchi2022what} applied traditional SFL to flaky tests, aiming to pinpoint the classes that cause the non-deterministic behavior of flaky tests.
Apart from using the tests' coverage behavior in the fashion of our \textit{single} baseline (RQ2), they also utilize the change history and static code metrics.
Their results indicate that this augmentation improves fault localization performance, ranking \SI{47}{\percent} of faulty classes in the top three most suspicious components, compared to \SI{39}{\percent} without.
Interestingly, the DStar formula yielded the worst performance, whereas this formula performed best in our experiments.
Unlike us, they do not consider the coverage behavior of both passing and failing executions of flaky tests (due to difficulties in reproducing them), which they also mention as a possible threat to validity.
Furthermore, SFFL works on the statement level instead of the class level and focuses only on coverage behavior, not other data sources such as change history and static code metrics.
The latter might, however, be a path worthwhile pursuing and a possible direction for future work.

Moran et al.\ proposed FlakyLoc~\cite{moranbarbon2020flakyloc}, a spectrum-based technique to identify environment factors leading to test flakiness.
They evaluated FlakyLoc on one flaky test where it was able to correctly identify that its intermittent failures were caused by a low screen resolution.
Unlike our approach, they deliberately alter the spectrum-based components (in their case environment factors) in a combinatorial fashion to identify the one responsible for the failure.
Coverage-based SFL (and thus SFFL) is not able to adopt this approach, since the spectrum-based components are code elements.

Ziftci and Cavalcanti~\cite{celalziftci2020de} proposed the Flakiness Debugger.
Like SFFL, it processes multiple coverage behaviors of flaky tests to determine the location of the flaky fault.
Unlike SFFL, it is not based on SFL, but on the custom \texttt{DIVERGENCE} algorithm that highlights the first divergence in the control flow between passing and failing executions.
Therefore, it also requires a more complex instrumentation, which does not only collect code coverage (which lines were covered), but full execution traces (the order in which the lines were executed).
Such instrumentation is tool-wise less supported and can have considerable impact on program behavior and therefore test flakiness.
They evaluated their approach on \num{83} fixed flaky tests, for which it was able to point to relevant code locations involved in flakiness with \SI{81.93}{\percent} accuracy.
In our evaluation, SFFL was able to identify the faults of \SfflDstarAcRankTenOrLessPercentageNod of flaky tests within the top ten most suspicious statements.
While being similar on a technical level, the Flakiness Debugger (and its evaluation) is targeted more towards aiding the manual debugging process, which is also why they preferred the \texttt{DIVERGENCE} algorithm over SFL.
SFFL can also be used for manual debugging, however, it can also build the foundation for automated repair of flaky tests, which Ziftci and Cavalcanti~\cite{celalziftci2020de} found to be in high demand by developers.

Ahmad et al.~\cite{ahmad2022identifying} introduced FLAKYPY, a tool that builds on the \texttt{DIVERGENCE} algorithm and aims to improve the localization of randomness-related flaky tests by taking the values of local variables and anomalies between them in different runs into account.
Applied to \num{26} projects taken from the same dataset as we use~\cite{gruber2021empirical},
 FLAKYPY was unable to reproduce the flaky behavior for \num{22} of them.
In \num{5} cases FLAKYPY was not able to find a potential root cause.
Unlike FLAKYPY, SFFL is not root cause specific, apart from targeting NOD flakiness.

\section{Conclusions}%
\label{sec:conclusions}

In this paper we introduced SFFL (Spectrum-based Flaky Fault Localization), an extension of SFL that specifically targets the debugging of flaky tests.
SFFL considers the fact that a test can have multiple coverage behaviors, and harnesses this property by computing the coverage intersection for flaky tests and the union for stable ones.
By doing so, SFFL remains mostly deterministic while achieving a substantial improvement over traditional SFL:
In our evaluation conducted on \NumNodTests non-order-dependent flaky tests it improved the average-case EXAM score in \ExamDstarAcImprovementOverFirstNumCasesGtzeroNod cases and by a mean of \ExamDstarAcImprovementOverFirstMeanNod.
In absolute terms, SFFL reached an average-case EXAM score of \ExamDstarAcSfflMeanNod and the faults of \SfflDstarAcRankTenOrLessPercentageNod of all flaky tests could be located within the top ten most suspicious statements.

A current limitation of our implementation of SFFL is that it is based
on a single-fault assumption, and we found that multiple fault
problems may even be amplified by SFFL's coverage processing. An
important aspect of future work will thus be to investigate extending
SFFL considering existing work on multi-fault SFL.
While there were no substantial differences between flaky tests of
different non-order-dependent root causes, a general limitation of
SFFL is that it targets non-order-dependent tests. Consequently,
future work could investigate generalizing SFFL to order-dependent
flaky tests, where the root cause for flakiness lies in the test code rather than the code under test.
Furthermore, it could try to apply SFFL in different contexts, such as other programming languages.

We hope that SFFL will both serve developers, making debugging flaky tests more convenient and efficient, and that it will inspire and be of use to future research, for example as a foundation for automated fixing or more sophisticated debugging techniques.
We make all data openly available~\cite{gruber2023SFFLdataset}.

\section*{Acknowledgements}\label{sec:acknowledgements}
We thank Christian Kasberger for his contributions to this paper in form of an early pre-study during his bachelor thesis.
We furthermore thank Florian Kroiss, Philipp Fink, and Ammar Harrat for manually labeling fault locations.

\balance
\bibliography{main.bbl}

\begin{thebibliography}{10}
\providecommand{\url}[1]{#1}
\csname url@samestyle\endcsname
\providecommand{\newblock}{\relax}
\providecommand{\bibinfo}[2]{#2}
\providecommand{\BIBentrySTDinterwordspacing}{\spaceskip=0pt\relax}
\providecommand{\BIBentryALTinterwordstretchfactor}{4}
\providecommand{\BIBentryALTinterwordspacing}{\spaceskip=\fontdimen2\font plus
\BIBentryALTinterwordstretchfactor\fontdimen3\font minus
  \fontdimen4\font\relax}
\providecommand{\BIBforeignlanguage}[2]{{%
\expandafter\ifx\csname l@#1\endcsname\relax
\typeout{** WARNING: IEEEtran.bst: No hyphenation pattern has been}%
\typeout{** loaded for the language `#1'. Using the pattern for}%
\typeout{** the default language instead.}%
\else
\language=\csname l@#1\endcsname
\fi
#2}}
\providecommand{\BIBdecl}{\relax}
\BIBdecl

\bibitem{gruber2022survey}
M.~Gruber and G.~Fraser, ``A survey on how test flakiness affects developers
  and what support they need to address it,'' in \emph{International Conference
  on Software Testing, Verification and Validation~(ICST)}, 2022, pp. 82--92.

\bibitem{parry2022surveying}
O.~Parry, G.~M. Kapfhammer, M.~Hilton, and P.~McMinn, ``Surveying the developer
  experience of flaky tests,'' in \emph{International Conference on Software
  Engineering: Software Engineering in Practice Track~(ICSE-SEIP)}, 2022, pp.
  253--262.

\bibitem{bell2018deflaker}
J.~Bell, O.~Legunsen, M.~Hilton, L.~Eloussi, T.~Yung, and D.~Marinov,
  ``{DeFlaker}: {Automatically} detecting flaky tests,'' in \emph{International
  Conference on Software Engineering~(ICSE)}, 2018, pp. 433--444.

\bibitem{lam2019idflakies}
W.~Lam, R.~Oei, A.~Shi, D.~Marinov, and T.~Xie, ``{iDFlakies}: A framework for
  detecting and partially classifying flaky tests,'' in \emph{International
  Conference on Software Testing, Verification and Validation~(ICST)}, 2019,
  pp. 312--322.

\bibitem{silva2020shake}
D.~Silva, L.~Teixeira, and M.~d’Amorim, ``Shake it! {Detecting} flaky tests
  caused by concurrency with {Shaker},'' in \emph{International Conference on
  Software Maintenance and Evolution~(ICSME)}, 2020, pp. 301--311.

\bibitem{pinto2020what}
G.~Pinto, B.~Miranda, S.~Dissanayake, M.~d'Amorim, C.~Treude, and A.~Bertolino,
  ``What is the vocabulary of flaky tests?'' in \emph{International Conference
  on Mining Software Repositories~(MSR)}, 2020, pp. 492--502.

\bibitem{alshammari2021flakeflagger}
A.~Alshammari, C.~Morris, M.~Hilton, and J.~Bell, ``{FlakeFlagger}:
  {Predicting} flakiness without rerunning tests,'' in \emph{International
  Conference on Software Engineering~(ICSE)}, 2021, pp. 1572--1584.

\bibitem{verdecchia2021know}
R.~Verdecchia, E.~Cruciani, B.~Miranda, and A.~Bertolino, ``Know you neighbor:
  Fast static prediction of test flakiness,'' \emph{IEEE Access}, pp.
  76\,119--76\,134, 2021.

\bibitem{fatima2022flakify}
S.~Fatima, T.~A. Ghaleb, and L.~Briand, ``Flakify: A black-box, language
  model-based predictor for flaky tests,'' \emph{{IEEE} Transactions on
  Software Engineering}, pp. 1--17, 2022.

\bibitem{qin2022peeler}
Y.~Qin, S.~Wang, K.~Liu, B.~Lin, H.~Wu, L.~Li, X.~Mao, and T.~F. D.~A.
  Bissyande, ``Peeler: Learning to effectively predict flakiness without
  running tests,'' in \emph{International Conference on Software Maintenance
  and Evolution~(ICSME)}, 2022, pp. 1--12.

\bibitem{parry2022evaluating}
O.~Parry, G.~M. Kapfhammer, M.~Hilton, and P.~McMinn, ``Evaluating features for
  machine learning detection of order-and non-order-dependent flaky tests,'' in
  \emph{International Conference on Software Testing, Verification and
  Validation~(ICST)}, 2022, pp. 93--104.

\bibitem{li2022evolution}
C.~Li and A.~Shi, ``Evolution-aware detection of order-dependent flaky tests,''
  in \emph{International Symposium on Software Testing and Analysis~(ISSTA)},
  2022, p. 114–125.

\bibitem{luo2014empirical}
Q.~Luo, F.~Hariri, L.~Eloussi, and D.~Marinov, ``An empirical analysis of flaky
  tests,'' in \emph{International Symposium on Foundations of Software
  Engineering~(FSE)}, 2014, pp. 643--653.

\bibitem{eck2019understanding}
M.~Eck, F.~Palomba, M.~Castelluccio, and A.~Bacchelli, ``Understanding flaky
  tests: The developer's perspective,'' in \emph{Joint Meeting of the European
  Software Engineering Conference and the Symposium on the Foundations of
  Software Engineering~(ESEC/FSE)}, 2019, pp. 830--840.

\bibitem{gruber2021empirical}
M.~Gruber, S.~Lukasczyk, F.~Kroi{\ss}, and G.~Fraser, ``An empirical study of
  flaky tests in {Python},'' in \emph{International Conference on Software
  Testing, Verification and Validation~(ICST)}, 2021, pp. 148--158.

\bibitem{gambi2018practical}
A.~Gambi, J.~Bell, and A.~Zeller, ``Practical test dependency detection,'' in
  \emph{International Conference on Software Testing, Verification and
  Validation~(ICST)}, 2018, pp. 1--11.

\bibitem{shi2019ifixflakies}
A.~Shi, W.~Lam, R.~Oei, T.~Xie, and D.~Marinov, ``{iFixFlakies}: {A} framework
  for automatically fixing order-dependent flaky tests,'' in \emph{Joint
  Meeting of the European Software Engineering Conference and the Symposium on
  the Foundations of Software Engineering~(ESEC/FSE)}, 2019, pp. 545--555.

\bibitem{wei2022preempting}
A.~Wei, P.~Yi, Z.~Li, T.~Xie, D.~Marinov, and W.~Lam, ``Preempting flaky tests
  via non-idempotent-outcome tests,'' in \emph{International Conference on
  Software Engineering~(ICSE)}, 2022, pp. 1730--1742.

\bibitem{wang2022ipflakies}
W.~L. Ruixin~Wang, Yang~Chen, ``{iPFlakies}: A framework for detecting and
  fixing python order-dependent flaky tests,'' in \emph{International
  Conference on Software Engineering: Companion Proceedings~(ICSE Companion)},
  2022, pp. 120--124.

\bibitem{gruber2023SFFLdataset}
\BIBentryALTinterwordspacing
M.~Gruber and G.~Fraser, ``Debugging flaky tests using spectrum-based fault
  localization [dataset],'' 2023. [Online]. Available:
  \url{https://doi.org/10.6084/m9.figshare.21901401}
\BIBentrySTDinterwordspacing

\bibitem{wong2016survey}
W.~E. Wong, R.~Gao, Y.~Li, R.~Abreu, and F.~Wotawa, ``A survey on software
  fault localization,'' \emph{IEEE Transactions on Software Engineering},
  vol.~42, no.~8, pp. 707--740, 2016.

\bibitem{gazzola2018automatic}
L.~Gazzola, D.~Micucci, and L.~Mariani, ``Automatic software repair: a
  survey,'' in \emph{International Conference on Software Engineering~(ICSE)},
  2018, p. 1219.

\bibitem{pearson2017evaluating}
S.~Pearson, J.~Campos, R.~Just, G.~Fraser, R.~Abreu, M.~D. Ernst, D.~Pang, and
  B.~Keller, ``Evaluating and improving fault localization,'' in
  \emph{International Conference on Software Engineering~(ICSE)}, 2017, pp.
  609--620.

\bibitem{wong2014dstar}
W.~E. Wong, V.~Debroy, R.~Gao, and Y.~Li, ``The {DStar} method for effective
  software fault localization,'' \emph{{IEEE} Transactions on Reliability},
  vol.~63, no.~1, pp. 290--308, 2014.

\bibitem{jones2002visualization}
J.~A. Jones, M.~J. Harrold, and J.~Stasko, ``Visualization of test information
  to assist fault localization,'' in \emph{International Conference on Software
  Engineering~(ICSE)}, 2002, pp. 467--477.

\bibitem{abreu2009practical}
R.~Abreu, P.~Zoeteweij, R.~Golsteijn, and A.~J.~C. van Gemund, ``A practical
  evaluation of spectrum-based fault localization,'' \emph{Journal of Systems
  and Software}, vol.~82, no.~11, pp. 1780--1792, 2009.

\bibitem{naish2011model}
L.~Naish, H.~J. Lee, and K.~Ramamohanarao, ``A model for spectra-based software
  diagnosis,'' \emph{{ACM} Transactions on Software Engineering and
  Methodology}, pp. 11:1--11:32, 2011.

\bibitem{abreu2009spectrum}
R.~Abreu, P.~Zoeteweij, and A.~J.~C. van Gemund, ``Spectrum-based multiple
  fault localization,'' in \emph{International Conference on Automated Software
  Engineering~(ASE)}, 2009, pp. 88--99.

\bibitem{cordy2022flakime}
M.~Cordy, R.~Rwemalika, A.~Franci, M.~Papadakis, and M.~Harman, ``{FlakiMe}:
  Laboratory-controlled test flakiness impact assessment,'' in
  \emph{International Conference on Software Engineering~(ICSE)}, 2022, pp.
  982--994.

\bibitem{qin2021impact}
Y.~Qin, S.~Wang, K.~Liu, X.~Mao, and T.~F. Bissyand{\'e}, ``On the impact of
  flaky tests in automated program repair,'' \emph{International Conference on
  Software Analysis, Evolution and Reengineering, {SANER}}, pp. 295--306, 2021.

\bibitem{vancsics2020simulating}
B.~Vancsics, T.~Gergely, and {\'A}.~Besz{\'e}des, ``Simulating the effect of
  test flakiness on fault localization effectiveness,'' in \emph{Workshop on
  Validation, Analysis and Evolution of Software Tests ({VST})}, 2020, pp.
  28--35.

\bibitem{habchi2022what}
S.~Habchi, G.~Haben, J.~Sohn, A.~Franci, M.~Papadakis, M.~Cordy, and Y.~L.
  Traon, ``What made this test flake? pinpointing classes responsible for test
  flakiness,'' in \emph{International Conference on Software Maintenance and
  Evolution~(ICSME)}, 2022, pp. 352--363.

\bibitem{moranbarbon2020flakyloc}
J.~Mor{\'a}n~Barb{\'o}n, C.~Augusto~Alonso, A.~Bertolino, C.~A.
  Riva~{\'A}lvarez, P.~J. Tuya~Gonz{\'a}lez \emph{et~al.}, ``Flakyloc:
  flakiness localization for reliable test suites in web applications,''
  \emph{Journal of Web Engineering, 2}, 2020.

\bibitem{celalziftci2020de}
D.~C. Celal~Ziftci, ``{De-Flake} your tests: Automatically locating root causes
  of flaky tests in code at {Google},'' in \emph{International Conference on
  Software Maintenance and Evolution~(ICSME)}, 2020, pp. 736--745.

\bibitem{hoffmann2011eclemma}
\BIBentryALTinterwordspacing
M.~R. Hoffmann, B.~Janiczak, and E.~Mandrikov, ``Eclemma-jacoco java code
  coverage library,'' 2011. [Online]. Available:
  \url{https://www.eclemma.org/jacoco/}
\BIBentrySTDinterwordspacing

\bibitem{pytestCov}
\BIBentryALTinterwordspacing
``pytest-cov: a pytest plugin that produces coverage reports.'' [Online].
  Available: \url{https://pytest-cov.readthedocs.io/en/v2.8.1/}
\BIBentrySTDinterwordspacing

\bibitem{micco2016flaky}
\BIBentryALTinterwordspacing
J.~Micco, ``Flaky tests at {Google} and how we mitigate them,'' 2016. [Online].
  Available:
  \url{https://testing.googleblog.com/2016/05/flaky-tests-at-google-and-how-we.html}
\BIBentrySTDinterwordspacing

\bibitem{elbaum2014techniques}
S.~Elbaum, G.~Rothermel, and J.~Penix, ``Techniques for improving regression
  testing in continuous integration development environments,'' in
  \emph{International Symposium on Foundations of Software Engineering~(FSE)},
  2014, pp. 235--245.

\bibitem{pytest}
\BIBentryALTinterwordspacing
``pytest.'' [Online]. Available: \url{https://docs.pytest.org/en/7.2.x/}
\BIBentrySTDinterwordspacing

\bibitem{pytestCovContext}
\BIBentryALTinterwordspacing
``pytest-cov contexts.'' [Online]. Available:
  \url{https://pytest-cov.readthedocs.io/en/v2.8.1/contexts.html}
\BIBentrySTDinterwordspacing

\bibitem{fleiss1971measuring}
J.~L. Fleiss, ``Measuring nominal scale agreement among many raters.''
  \emph{Psychological bulletin}, p. 378, 1971.

\bibitem{regier2013dsm}
D.~A. Regier, W.~E. Narrow, D.~E. Clarke, H.~C. Kraemer, S.~J. Kuramoto, E.~A.
  Kuhl, and D.~J. Kupfer, ``Dsm-5 field trials in the united states and canada,
  part ii: test-retest reliability of selected categorical diagnoses,''
  \emph{American journal of psychiatry}, pp. 59--70, 2013.

\bibitem{wong2008crosstab}
W.~E. Wong, T.~Wei, Y.~Qi, and L.~Zhao, ``A crosstab-based statistical method
  for effective fault localization,'' in \emph{International Conference on
  Software Testing, Verification and Validation~(ICST)}, 2008, pp. 42--51.

\bibitem{adanial_cloc}
\BIBentryALTinterwordspacing
A.~Danial, ``cloc: v1.92,'' 2021. [Online]. Available:
  \url{https://doi.org/10.5281/zenodo.5760077}
\BIBentrySTDinterwordspacing

\bibitem{wilcoxon1945individual}
F.~Wilcoxon, ``Individual comparisons by ranking methods,'' \emph{Biometrics
  Bulletin}, pp. 80--83, 1945.

\bibitem{mann1947test}
H.~B. Mann and D.~R. Whitney, ``On a test of whether one of two random
  variables is stochastically larger than the other,'' \emph{The annals of
  mathematical statistics}, pp. 50--60, 1947.

\bibitem{vargha2000critique}
A.~Vargha and H.~D. Delaney, ``A critique and improvement of the cl common
  language effect size statistics of mcgraw and wong,'' \emph{Journal of
  Educational and Behavioral Statistics}, pp. 101--132, 2000.

\bibitem{ahmad2022identifying}
A.~Ahmad, E.~N. Held, O.~Leifler, and K.~Sandahl, ``Identifying randomness
  related flaky tests through divergence and execution tracing,'' in
  \emph{International Conference on Software Testing, Verification and
  Validation~(ICST)}, 2022, pp. 293--300.

\bibitem{parnin2011are}
C.~Parnin and A.~Orso, ``Are automated debugging techniques actually helping
  programmers?'' in \emph{International Symposium on Software Testing and
  Analysis~(ISSTA)}, 2011, pp. 199--209.

\bibitem{maciver2019hypothesis}
D.~Maciver and Z.~Hatfield{-}Dodds, ``Hypothesis: {A} new approach to
  property-based testing,'' \emph{J. Open Source Softw.}, vol.~4, no.~43, p.
  1891, 2019.

\bibitem{just2014defects4j}
R.~Just, D.~Jalali, and M.~D. Ernst, ``Defects4j: A database of existing faults
  to enable controlled testing studies for java programs,'' in
  \emph{International Symposium on Software Testing and Analysis~(ISSTA)},
  2014, pp. 437--440.

\bibitem{martinez2017automatic}
M.~Martinez, T.~Durieux, R.~Sommerard, J.~Xuan, and M.~Monperrus, ``Automatic
  repair of real bugs in java: A large-scale experiment on the defects4j
  dataset,'' \emph{Empirical Software Engineering}, vol.~22, no.~4, pp.
  1936--1964, 2017.

\end{thebibliography}

\vspace{5pt}
\centering
All online resources accessed on 2023--01--16.

\end{document}